\documentclass{aa}  
\usepackage[varg]{txfonts}
\usepackage{graphicx}
\usepackage{amsmath}
\usepackage{makecell,booktabs}
\usepackage{orcidlink}
\usepackage{placeins}
\usepackage{natbib}
\usepackage{hyperref}
\hypersetup{
    colorlinks,
    citecolor=blue,
    linkcolor=blue,
    filecolor=magenta,      
    urlcolor=cyan,
    pdftitle={M.Birney et al.: kinematical study of the launching region of the HH 46/47 outflow},
    pdfpagemode=FullScreen,
    }
\bibpunct{(}{)}{;}{a}{}{,}
\begin{document} 

   \title{A kinematical study of the launching region of the blueshifted HH 46/47 outflow with SINFONI K-band observations\thanks{Based on observations collected with the VLT operated by the European Southern Observatory under ESO program ID 0102.C-0615(A).}}
    %\thanks{test}
   %\subtitle{I. Kinematics and morphology}

   \author{M. Birney\inst{\ref{inst1}}\orcidlink{0000-0002-6627-9863}\thanks{Corresponding author; \texttt{matthew.birney@mu.ie}},
   C. Dougados\inst{\ref{inst2}}\orcidlink{0000-0001-6660-936X}, 
   E. T. Whelan\inst{\ref{inst1}}\orcidlink{0000-0002-3741-9353}, 
   B. Nisini\inst{\ref{inst3}}\orcidlink{0000-0002-9190-0113}, 
   S. Cabrit\inst{\ref{inst2}, \ref{inst4}}\orcidlink{0000-0002-1593-3693} 
   \and 
   Y. Zhang\inst{\ref{inst5}}\orcidlink{0000-0001-7511-0034} \fnmsep
   %\thanks{test}
          }

   \institute{Department of Physics, Maynooth University, Maynooth, Co. Kildare, Ireland \label{inst1} \\
    \and
    Univ. Grenoble Alpes, CNRS, IPAG, 38000 Grenoble, France \label{inst2}\
    \and
    INAF – Osservatorio Astronomico di Roma, Via di Frascati 3300074, Monte Porzio Catone, Italy \label{inst3}\
    \and
    PSL University, Sorbonne Université, Observatoire de Paris, LERMA, CNRS UMR 8112, 75014 Paris, France\label{inst4}\
    \and
    Department of Astronomy, Shanghai Jiao Tong University, 800 Dongchuan Rd., Minhang, Shanghai 200240, China\label{inst5}\
    %\thanks{The university of heaven temporarily does not accept e-mails}
             }

   \date{Received 9 July 2024 / Accepted 2 Novemeber 2024}

% \abstract{}{}{}{}{} 
% 5 {} token are mandatory
 
  \abstract
    {Studying outflows is important, as they may significantly contribute to angular momentum removal from a star-disc system and thus affect disc evolution and planet formation.}
   {To investigate the different outflow components, including the collimated jet, wide-angled molecular outflow, and outflow cavity, of the Class I HH 46/47 outflow system, we focused on their kinematics.}
   {We present near-infrared (NIR) K-band integral field observations of the blueshifted HH 46/47 outflow base obtained using VLT/SINFONI with an angular resolution of 0\farcs81. Our analysis focuses on [\ion{Fe}{ii}], H$_2$ 1–0 S(1), and Br-$\gamma$ emission. We employed a wavelength recalibration technique based on OH telluric lines in order to probe the kinematics of the wide-angled flow with an accuracy of $\sim$ 1~km~s$^{-1}$ - 3~km~s$^{-1}$.}
   {We confirmed a velocity gradient of $\sim$ 10 km~s$^{-1}$ transverse to the outflow direction in the wide-angled H$_2$ outflow cavity. We find that the H$_2$ cavity peaks at radial velocities of $\sim$ $-$15 km~s$^{-1}$ to $-$30 km~s$^{-1}$, and that the atomic jet peaks at $v_{\mathrm{rad}}$ $\sim$ $-$210 km~s$^{-1}$. The outflow exhibits a layered structure: The high-velocity [\ion{Fe}{ii}] and Br-$\gamma$ jet is surrounded by a wide-angled H$_2$ outflow cavity that is in turn nested within the continuum emission and CO molecular outflow. The continuum emission and H$_2$ outflow cavity are asymmetric with respect to the jet axis.}
   {We propose that the origin of the asymmetries and the velocity gradient detected in the wide-angled H$_2$ cavity is due to a wide-angled outflow or successive jet bowshocks expanding into an inhomogeneous ambient medium or the presence of a secondary outflow. We eliminated outflow rotation as an exclusive origin of this velocity gradient due to large specific angular momenta values, $J(r)$ $\approx$ 3000 - 4000 km~s$^{-1} \,$au, calculated from 1\arcsec\ to 2\arcsec\ along the outflow and the fact that the sense of apparent rotation we detected is opposite to that of the CO envelope emission. The observations reveal the complexities inherent in outflow systems and the risk of attributing transverse velocity gradients solely to rotation.}


%\titlerunning{short title}
%\authorrunning{short authors}
\titlerunning{A kinematical study of the launching region of the HH 46/47 outflow}
\authorrunning{M. Birney et al.}
\maketitle
%
%-------------------------------------------------------------------

\section{Introduction} \label{Introduction}

% \begin{enumerate}
% \item Add Sylvie as an author 
% \item Once you define an acronym make sure you use it for the rest of the paper e.g. PV 
% \item Sometime you use figure and sometimes Figure - It should be capitalised I think
% \item In section 3 first paragraph where you say what emission lines you use you could sau H$_{2}$ at 2.122 $\mu$m (hereafter H$_{2}$) and then just use H$_{2}$ for the rest of the paper. The same for the [FeII] line you use and the Br-gamma. Youn jump back and forth throughout the manuscript between the shorthand for the line and the full line name e.g Br~$\gamma$ 2.166 and Br~$\gamma$ and again this affects the flow of the paper. 
% \item Avoid repeating words in a sentence.
% \item There is a lot of repetition for example you say several times throughout the manuscript that the jet bends towards the Northern arm. You should think of the paper as telling a story and removing this kind of repetition will make it flow better.
% \item Introduction could be more tailored to you results and discussion - its a little general in parts now. 
% \item some paragraphs are very small and could be merged. 
% \item I think there is a lot of overlap between section 2.7 and the start of the results. I have done some re-organsisation
% \item we need to be very clear from the start of the results about what the 0 point is and how we measure that.
% \item When you say jet axis in relation to asymmetry about the jet axis what exactly do you mean - do you mean the line perpendicular to the measured disk PA. Do we know the PA? Lets talk about this more
% \end{enumerate}

Outflows (jets and winds) from young stellar objects (YSOs) are ubiquitous, but their origin is a major open question in star formation \citep{Frank2014, Pascucci2023}. They are launched very early in the star formation process perpendicularly to the accretion disc surrounding the YSO and can persist for several million years \citep{Whelan2014}. Ejection in YSOs is intrinsically linked with accretion, and the outflows interact with the surrounding envelope of quiescent material, creating shocked regions \citep{Hartigan2003, Arce2007}. Studying YSO outflows is important, as they may significantly contribute to the angular momentum removal from the star-disc system, which is an unsolved dilemma in star formation \citep{Pudritz2019}. Angular momentum transport plays a central role in the star formation process, as the initial angular momentum present in dense star forming cores needs to be reduced by several orders of magnitude in order to permit a star or a multiple star system to form \citep{Belloche2013}. Outflows also influence disc evolution and likely affect planet formation and migration, as they may interact with the planet forming regions of the disc \citep{Pascucci2023}. 

Observing protostars is essential to gaining insights on the most active period of accretion and ejection from a forming star, and Class 0/I low-mass sources have been much studied to date \citep{Ray&Ferreira2021}. However, protostars are challenging to observe due to their heavily embedded nature. Their dusty and gaseous envelopes cause strong extinction, making it difficult to probe the inner regions of the discs where accretion occurs and where the earliest instances of planet formation are beginning. One can, however, learn about the ejection, and indirectly the inner disc properties, of these YSOs by studying their powerful outflows once the most suitable tracers have been selected. Protostellar outflows from Class 0/I YSOs consist of high-velocity ($v$ $\sim$ 100 km~s$^{-1}$ - 400 km~s$^{-1}$) collimated jets launched from the innermost region of the discs, lower-velocity ($v$ $\sim$ 1 km~s$^{-1}$ - 30 km~s$^{-1}$) wide-angled winds launched from larger disc radii \citep{Frank2014, Pascucci2023}, and entrained material from the surrounding envelope \citep{Arce2013}. Lower-velocity wider-outflow components surrounding the bases of class I jets were first discovered by \cite{Davis2002} using near-infrared (NIR) narrow-band H$_2$ imaging. Cavities, that are defined as hollow low-density regions created by the outflow, surrounding atomic jets are well documented phenomena \citep{Habel2021}, and Class I protostars with such cavities are regularly seen. The origin of these outflow cavities is currently debated. They may be directly or partly tracing intrinsic wide-angled disc winds originating from the disc surface \citep{deValon2022, LopezVazquez2024}, either by magnetic (magnetohydrodynamic) processes or thermal (photo-evaporative) processes. The cavities may be tracing an interaction region between some kind of wide-angled disc wind and the surrounding medium, be it static or an infalling envelope. They may also be produced by the stacking of successive jet bowshocks from a collimated, variable inner jet propagating into a density-stratified surrounding medium \citep{Rabenanahary2022}. 

Magnetohydrodynamic (MHD) processes in the rotating star-disc system, where magnetic field lines anchored in the accretion disc magneto-centrifugally launch material from the disc plane, are theorised to give rise to outflows \citep{Ray&Ferreira2021}. Current models predict that MHD outflows emanate from a range of disc radii (the disc wind model) or from a single narrow radial region \citep[the X-wind model;][]{Ferreira1997, Bai2016}. In the disc wind model \citep{Pudritz&Norman1983, Konigl&Pudritz2000, Pudritz2007}, the high-velocity collimated jet component is launched from the innermost regions of the disc, whilst the launching region of the lower-velocity wide-angled outflow component is more radially extended along the disc and distributed over a larger range of disc radii. This is significant for both disc evolution and later planet formation, as wide-angled winds may substantially affect these processes. An alternative model for launching an outflow magneto-centrifugally is the X-wind model \citep{Shu2000}. In this model, the outflow is launched from a narrow radial region at the interface region between the inner disc edge and the stellar magnetosphere, the so-called co-rotation radius. High-velocity collimated jets and slower moving wide-angled winds are also predicted with this model \citep{Shang2020}. However, in contrast with the disc wind model, the wide-angled wind component in the X-wind model flows at much higher velocities, as all of the magnetic outflow streamlines originate at the same disc radius. 

Disc wind models predict a greater fraction of angular momentum removal from the star-disc system, as the outflow driving region extends further radially along the disc from the driving source, and specific angular momentum increases with disc radii \citep{Anderson2003, Ferreira2006, Pudritz2019, Lee2021}. Therefore, measurements of outflow rotation that lead to estimates of specific angular momentum are critical for constraining the two MHD models, and to date they have tended to support the disc wind model \citep{Coffey2017}. Wide-angled winds can offer information about angular momentum transport. As the lateral component of these wide-angled winds can more readily be spatially resolved, obtaining outflow rotation measurements is more promptly achievable \citep{Tabone2017, Zhang2018}. Measuring outflow rotation signatures using only the high-velocity jet counterparts is more challenging and requires high-angular resolution observations \citep{Coffey2004, Coffey2007, Lee2017}. 

Outflow emission arises primarily from the cooling regions of the shocks generated when higher-velocity material catches up with and impacts material ahead of it \citep{Tsukamoto2023}. While outflows are observed in all wavelength regimes, Class 0/I outflows are best observed with IR and sub-millimetre observations, as these wavelengths allow one to peer through the dusty envelope \citep{Yang2022, Harsono2023, Beuther2023, Ray2023, Nisini2024, Delabrosse2024}. In particular, these outflows can be studied using IR Fe forbidden and H$_2$ ro-vibrational emission lines, as these lines emit brightly near their bases, and interestingly these two species trace different outflow components \citep{Davis2011}. [\ion{Fe}{ii}] emission traces hot, dense, and partially ionised high-velocity gas ($T$ $\sim$ 10,000 K; $n_{\mathrm{e}}$ $\sim$ 10$^5$ cm$^{-3}$; $v$ $\sim$ 100 km~s$^{-1}$ - 400 km~s$^{-1}$; \citealt{Nisini2002, Davis2003}), while H$_2$ traces a molecular hydrogen emission line (MHEL) region associated with lower excitation, shocked molecular gas, or UV fluoresced gas at much lower velocities ($T$ $\sim$ 2,000 K; $n_{\mathrm{e}}$ $\sim$ 10$^3$ cm$^{-3}$; $v$ $\sim$ 10 km~s$^{-1}$ - 50 km~s$^{-1}$; \citealt{Davis2002, CarattiOGaratti2006, Frank2014}). Therefore, [\ion{Fe}{ii}] emission is a well-established atomic jet tracer \citep{Davis2003}, and shocked H$_2$ emission is an important means of tracing molecular winds and cavities \citep{Nisini2024}. The CO rotational emission lines at sub-millimetre wavelengths are also found in outflows from Class 0/I YSOs. The component of the outflow that CO likely maps is the resulting ambient gas after it has been swept up and entrained by some wide-angled wind and later left to cool \citep{Arce2013}. However, in some cases it is thought to trace the disc wind itself \citep{Louvet2018, deValon2022}. The CO emission traces molecular gas at low velocities also and is cooler than the H$_2$ emission ($T$ $\sim$ 100 K; $n_{\mathrm{e}}$ $\sim$ 10$^3$ cm$^{-3}$). Therefore, CO emission complements snapshots of current shock interactions obtained with H$_2$ and [\ion{Fe}{ii]} observations by providing insight into the outflow history of the protostar \citep{Richer2000}.

In this work, we present Spectrograph for INtegral Field Observations in the Near Infrared (SINFONI) integral field spectroscopy observations of the HH 46/47 outflow system in the K-band (1.95 - 2.45 $\mu$m) with a resolving power of R = 4000 (Fig. \ref{Morphologyfig}). Located in the Bok Globule ESO 210-6A in the Gum nebula, HH 46/47 is an excellent prototype system for isolated star formation (see Sect. \ref{HH4647} for more details on the target). The field of view (FOV) chosen was 8\farcs0 $\times$ 8\farcs0, resulting in an observation of the base of the blueshifted outflow. These observations allowed us to investigate both the base of the fast axial jet and the wide-angled outflow component in NIR [\ion{Fe}{ii}] and H$_2$ emission and to subsequently make a comparison with previous CO sub-millimetre observations. In particular, we wished to use a wavelength recalibration technique based on OH lines to search for a rotation signature in the HH 46/47 outflow. The target, observations, data reduction, and data calibration are discussed in Sect. \ref{Sec: target, obs. data red.}. In Sect. \ref{sec: results}, we present the results of our observations, namely the morphologies and kinematics of the continuum emission, the wide-angled H$_2$ outflow cavity, and the base of the atomic jet. We also present a velocity gradient transverse to the outflow axis observed in the H$_2$ outflow cavity. In Sect. \ref{sec: Discussion}, we discuss our results. Included in our discussion is the observed layering of different emission tracers and the origin of the velocity gradient associated with the H$_2$ outflow cavity. We summarise our conclusions in Sect. \ref{sec: Conclusions}.

\section{Target, observations, data reduction, and calibration} \label{Sec: target, obs. data red.}
\subsection{HH 46/47} \label{HH4647}

\begin{figure}
   \centering
   \includegraphics[width=8.8cm, trim= 0cm 0cm 0cm 0cm, clip=true]{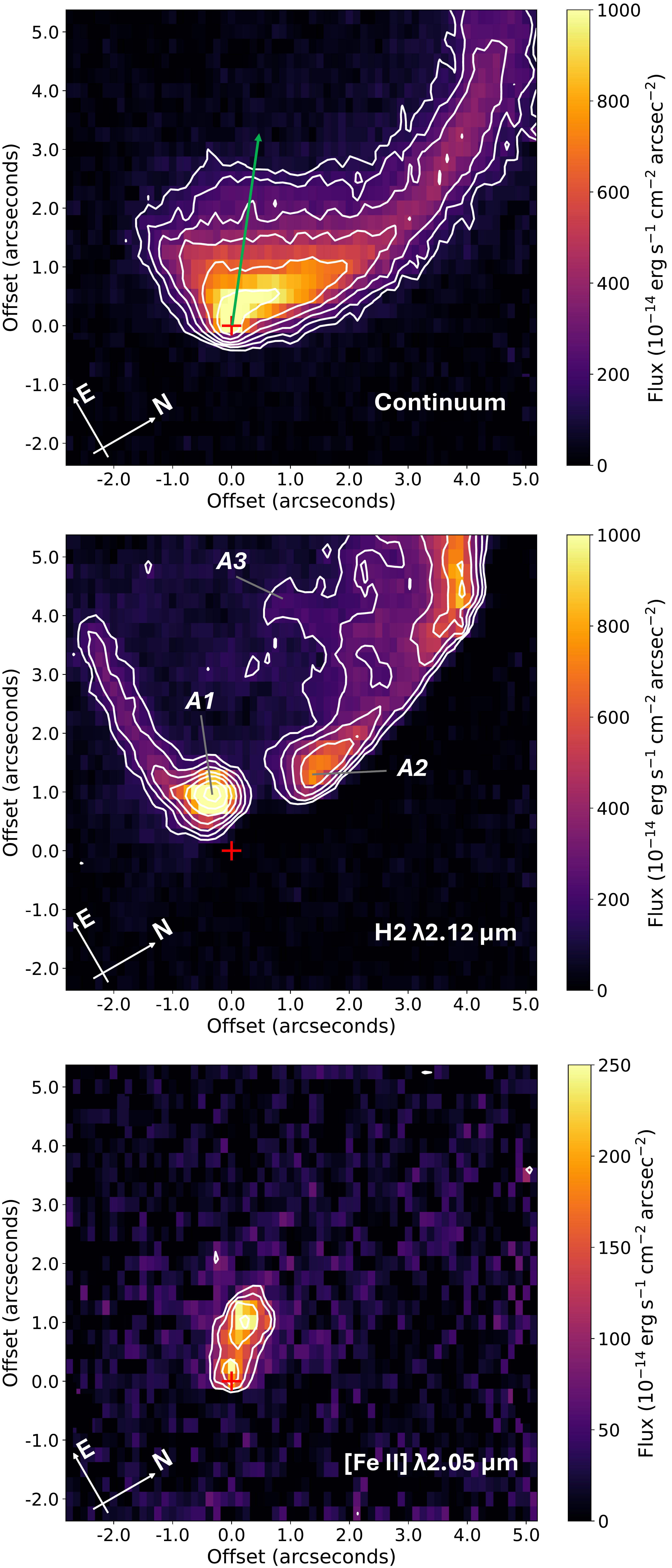}
   \caption{Integrated intensity maps displaying the morphologies of the outflow in different emission tracers. The source position is marked with a red plus. Contour levels for all begin at 3$\sigma$ of the background emission and increase by factors of 1.5. The vertical axis is aligned with centre axis of CO outflow (PA = 60\textdegree). Top: Continuum emission displaying a wide, approximately parabolic shape. Average jet PA of 52.3\textdegree\ indicated by a green arrow. Centre: Continuum subtracted and scattered light subtracted, integrated H$_2$ emission displaying a V-shaped outflow cavity. Bottom: Continuum subtracted, integrated [\ion{Fe}{ii}] emission displaying collimated atomic jet emission.}
              \label{Morphologyfig}%
\end{figure}

Discovered by \citet{Schwartz1977}, HH 46/47 is a well-studied complex of Herbig-Haro (HH) objects, a series of internal working surfaces or shock fronts in a larger outflow. The driving source of the outflow system (HH 46 IRS, HH 47 IRS, IRAS 08242-5050, 2MASS J08254384-5100326) is an embedded (A$_V$ > 35), low-mass, Class I YSO binary system of $\sim$ 0\farcs26 separation or 117 au \citep{Reipurth2000, Nisini2024}, taking a distance to the system of 450~pc \citep{GrahamHeyer1989}. The source has a mass of approximately 1.2 M$_{\odot}$ and bolometric luminosity of < 15 L$_{\odot}$ \citep{Antoniucci2008}, and is located at the northeastern edge of the Bok globule, ESO 210-6A, in the Gum nebula. The bipolar outflow system consists of wide-angled low-velocity molecular outflows accompanied by highly collimated high-velocity mainly atomic jets. The blueshifted side of the jet emanates northeasterly from the Bok Globule, making it easily visible at optical wavelengths \citep{Heathcote1996, Hartigan2011}. The redshifted side of the jet penetrates southwesterly, deep into the cloud, making it, and the driving source itself, visible at only infrared wavelengths \citep{Eisloffel1994, Reipurth2000, Noriega-Crespo2004, Erkal2021}.

The blueshifted jet displays a screw-like structure that spirals counter-clockwise towards the observer with an angle of inclination of 37\textdegree\ with respect to the plane of the sky \citep{Hartigan2005}. Proper motions at the base of the jet are approximately 270 km~s$^{-1}$ and increases to 300 km~s$^{-1}$ \citep{Hartigan2005, Erkal2021}. Wide-field images reveal that HH 46/47 extends at least 2.6 pc from the driving source, setting a lower limit to the dynamical age of the outflow of $10^4$ – $10^5$ years old \citep{Stanke1999}. The jets display a prominent ‘wiggling'. It was postulated that the orbital motion of an unresolved tertiary component caused the meandering of the jet \citep{Reipurth2000}, as the separation of the binary system was too large to explain the short wiggling period of $\sim$ 200 yr of the jet seen at large scales. The jet also displays a highly structured morphology, with knots seen along the outflow implying episodic ejection of material from the driving source. 

The wide-angled molecular outflows have been well studied using ALMA CO (1-0) observations \citep{Arce2013, Zhang2016} and CO (2-1) observations \citep{Zhang2019}. The low-velocity  wide-angled CO outflow is highly asymmetric, with the redshifted lobe extending four times further than blueshifted lobe, a lack of dense material in the path of the blueshifted flow is responsible for this. This wide-angled component displays multiple shell-like structures and these shells trace the ejection variability just like the collimated component of the outflow \citep{Zhang2019}. The variability timescales of the multiple outbursts seen in the wide-angle wind is consistent with episodic knots along the jet of HH 46/47. This suggests that the multiple shell structure seen in wide-angle component may arise from the same high accretion rate episodes that is reflected in the knots in the jets.

H$_2$ emission was first detected in the HH 46/47 system by \cite{Schwartz1983} with the identification of six fluorescent UV emission lines produced by Ly-$\alpha$ pumping of H$_2$. These molecules had to be first warmed to the second vibrational state of the electronic ground state. Later, NIR H$_2$ quadrupole emission in HH 46/47 was detected by \cite{Wilking1990}. \cite{Eisloffel1994} first presented the H$_2$ morphology of the HH 46/47 system using the IRAC2 instrument on the ESO/MPI 2.2~m telescope. \cite{Noriega-Crespo2004} imaged the system in the near to mid-infrared using the Spitzer Space Telescope revealing the optically invisible redshifted jet propagating southwesterly exhibiting a loop-like morphology, with both outflow lobes being narrower than those observed in CO emission. \cite{Garcia-Lopez2010} used long-slit NIR spectroscopic observations of the system using ISAAC and investigated the kinematics of the atomic and molecular emission, namely H$_2$ and [\ion{Fe}{ii}] emission, along the jet. 

\cite{Erkal2021} mapped the HH 46/47 system with HST with, at the time, unprecedented detail and recently, the HH 46/47 driving source and outflow were mapped at higher angular resolution of $\sim$ 0\farcs2 with JWST using the NIRSpec IFU and MIRI MRS as part of the PROJECT-J program (ID 1706, P: B. Nisini; \citealt{Nisini2024}). These observations provide unprecedented detail of the structure of the jet, the associated molecular outflow, and the cavity. Emission in H$_2$ reveals a complex molecular outflow, where the bright outflow cavity, expanding molecular shells and jet-driven bow-shocks interact with, and are influenced by ambient conditions. The superior spectral resolution provided by our SINFONI observations with respect to JWST (R = 4000 vs 2700), allows the outflow kinematics to be studied with unprecedented detail.

\subsection{VLT SINFONI} \label{VLT_SINFONI}
Observations of the HH 46/47 outflow were obtained with SINFONI on 27 November 2018 under ESO Program ID 0102.C-0615(A) (PI: C. Dougados). The NIR observations were obtained using the K-band grating (1.95 - 2.45 $\mu$m) with a resolving power of R = 4000. The FOV chosen was 8\farcs0 $\times$ 8\farcs0 providing a spatial sampling of 0\farcs125 $\times$ 0\farcs250 \citep{Eisenhauer2003}. This FOV was chosen as it was well suited to encompass the base of the outflow. The detector was rotated 60\textdegree\ so that the vertical axis of the detector was parallel to the approximate CO outflow position angle \citep[PA;][]{Zhang2019}. The observations consisted of ten exposures of 150~s each providing an on target exposure time of 1500~s. The average seeing of the science observation was 0\farcs81. It was not possible to use the instruments adaptive optics (AO) correction due to a lack of a suitable bright point source nearby. Three exposures of 300 s were also taken for sky exposures. Two standard stars, HD 67954 (HIP 039774) and HD 73105 (HIP 042038), were also observed for flux calibration and telluric correction. The goal of the observations was to extract the 2D spatial and kinematic distribution of atomic and H$_2$ emission lines at the base of the blueshifted outflow cavity. 

\subsection{Data reduction} \label{datareduction}
The data was reduced using ESO's SINFONI data reduction pipeline (Version 3.3.0) and run with ESOReflex \citep{Freudling2013}. Standard calibration files were used. The data cube was corrected for bad pixels, flat-fielding, bias and dark signal, and sky emission, and a first order wavelength calibration was done using arc lamp wavelength solutions. All emission line velocities discussed throughout are corrected for the Barycentric Earth radial velocity (BERV) and the Local Standard of Rest (LSR) velocity of the cloud that the outflow emanates from, which is +5.3 km~s$^{-1}$ \citep{Vankempen2009}. The kinematical measurements presented here are therefore expressed with respect to the frame of the central source.

\subsection{Wavelength recalibration} \label{WavelengthRecal}
The spectral resolution of SINFONI in the K-band is approximately R = 4000 or 2.45 Å, which translates to a velocity resolution of 75 km~s$^{-1}$. The spectral sampling size in terms of velocity is 38 km~s$^{-1}$ at 1.95 $\mu$m and 30 km~s$^{-1}$ at 2.45 $\mu$m. The wavelength calibration was performed using the standard procedure with the SINFONI data reduction pipeline. Wavelength calibration errors of approximately $\pm$ 30 km~s$^{-1}$ were observed, by fitting OH telluric emission lines from Earth's upper atmosphere, in the final data cube after data reduction using the SINFONI pipeline, which are not optimal for kinematical analysis. In order to increase the velocity calibration accuracy, a more rigorous wavelength calibration technique was employed. The data was run through the data reduction pipeline again, but the sky subtraction step was omitted from the process. This resulted in a data cube that contained the OH telluric emission lines. As these lines have a radial velocity of 0 km~s$^{-1}$ with respect to Earth's surface, they serve as a valuable tool for wavelength calibration. A 2D wavelength correction map could be produced using these OH lines as a velocity reference point, accounting for possible wavelength drifts across the FOV of the observation.

 % However, this velocity resolution was not optimal for the kinematical analyses we planned to do as typical radial velocities of molecular outflows traced in H$_2$ are between 10 and 50 km~s$^{-1}$.
 
A cross-correlation method was employed to produce this 2D wavelength correction array by comparing the OH emission lines present in the science cube spectra with a theoretical night-sky spectrum. The cross-correlation routine used was from the Toolkit for Exoplanet deTection and chaRacterization with IfS \citep[TExTRIS;][]{Bonnefoy2014, Petrus2021, Palma-Bifani2023, Demars2023}. The ESO SKYCALC tool \citep{Noll2012, Jones2013} was used to generate a synthetic night-sky spectrum that matched the night of observation and observatory location. The synthetic spectrum had a spectral resolution of R = 100,000 and was convolved with a Gaussian in order to downscale the spectral resolution to match that of the SINFONI K-band observations. 

The cross-correlation procedure compares the spectral positions of the OH lines in the science cube with the theoretical positions in the synthetic night sky spectrum, and calculates the wavelength offset between the two around a desired spectral region. The cross-correlation method was run in a spectral region centred on the H$_{2}$ 2.12 $\mu$m emission line with care taken to avoid large telluric absorption bands within the spectral window. A 2D wavelength correction map was produced, where each pixel in the map corresponded to a wavelength offset between the OH lines in the science spectra and theoretical spectrum. A linear trend was observed in the each of the horizontal rows of pixels in the 2D map. This trend is in accordance with the orientation of the horizontal slicing mirrors of the beam splitter present in the SINFONI integral field unit (IFU). Each horizontal row of pixels in the 2D wavelength correction map was isolated and a linear fit was applied to the data points. Each linear fit was compiled into a 2D wavelength correction map. This wavelength correction map is shown in the central panel of Fig. \ref{OHmap}. This method of precise wavelength calibration was also performed on SINFONI observations in \cite{Delabrosse2024}.

This final wavelength correction map was applied to the SINFONI data cube, effectively obtaining a more accurate wavelength calibration than that offered with the standard arc lamp wavelength calibration. The relative uncertainty associated with this wavelength correction map, calculated from the standard deviations of the residuals of the horizontal linear fits, is $\pm$ 0.55 km~s$^{-1}$. The absolute error associated with the map is estimated to be $\pm$ 0.62 km~s$^{-1}$, calculated by computing the wavelength correction map in several narrow wavelength intervals throughout the spectral axis, allowing us to estimate the variation in wavelength (velocity) correction values as a function of the central wavelength. Summing the absolute and relative error in quadrature we retrieve the total uncertainty of the wavelength calibration procedure to be $\pm$ 0.83 km~s$^{-1}$. We note that this value is not the final uncertainty on individual line centroid velocities. 

There are also systematic errors that arise when using this wavelength calibration method that must be considered, as a single average wavelength correction value is derived for each spatial-pixel (spaxel) over a wavelength range. \cite{Delabrosse2024} noticed when applying the 2D wavelength correction map to single OH lines in the vicinity of the H$_{2}$ 2.12 $\mu$m emission line, there existed systematic offsets from the OH line's actual wavelength. These systematic uncertainties produce absolute errors of the order of 0.4 km~s$^{-1}$, retrieved from Gaussian centroid offsets from the OH lines' known wavelengths, and relative uncertainties of the order of 0.9 km~s$^{-1}$, retrieved from the standard deviation of wavelength correction values in the 2D wavelength correction map applied to the OH lines. We adopt the same values as we find similar results, as expected using the same wavelength calibration method. We therefore estimate a 1$\sigma$ uncertainty associated with the wavelength calibration of $\pm$ 1.0 km~s$^{-1}$.

\subsection{Uneven slit effect correction} \label{sec: Unevenslit}
The uneven slit effect is an instrumental limitation associated with spectrometers due to the non-uniform illumination of the slit, or an illumination gradient within the slit. In the case of IFUs, the slicing mirrors, that individually behave like long-slit spectrographs, are illuminated non-uniformly, introducing a spurious wavelength shift in the spectral axis between on-axis and off-axis light. Incoming light that enters the slicing mirror or slitlet at a point above the slitlet centre will have an opposite wavelength bias when compared with incoming light entering the slit at a point below the slitlet centre \citep{Whelan&Garcia2008}. In the case of these observations, the effect arises due to the slope of the PSF creating a non-uniform brightness gradient within the slit. The resulting spurious wavelength shift depends on how strong the local brightness gradient is across the slit.  

The generation of a further 2D wavelength correction map was necessary in order to correct for these illusory wavelength shifts due to uneven slit illumination. In our case, we endeavoured to generate a wavelength correction map for the H$_{2}$ 2.12 $\mu$m emission line, as we required optimal confidence in the radial velocity measurements of the molecular outflow. In the case of SINFONI, the horizontal slicing mirrors can introduce spurious wavelength shifts up to a few kilometres per second \citep{Agra-Amboage2014}. The wavelength correction map due to uneven slit illumination was generated following the formulation developed by \cite{Marconi2003}. This method is used successfully in \cite{Erkal2021_b} and \cite{Agra-Amboage2014}. The correction for the uneven slit effect is explained further in Appendix \ref{AppSect: Uneven slit}. Figure \ref{unevenslitmap} displays the resulting 2D wavelength correction map to amend for uneven slit illumination, generated for the H$_{2}$ 2.12 $\mu$m emission line, was applied to the data cube around this emission line of interest. 

\subsection{Flux calibration and telluric correction} \label{sec: Fluxcal}
Flux calibration was performed on the SINFONI data cube using the two standard star observations that were part of the same observation run. The standard star observations, HD 67954 (HIP 39774) \& HD 73105 (HIP 42038), were reduced with the SINFONI data reduction pipeline in the same fashion as the science cube. Apertures, slightly larger in size than the seeing disc radius, were placed over the standard star's stellar positions and integrated to obtain the total flux with a high S/N. Annuli were used to estimate the contribution from the local background, which was subsequently subtracted from the aperture integration. Integrated stellar spectra were extracted using this method. The standard star's known magnitudes in J, H and K were obtained from the 2MASS catalogue \citep{Skrutskie2006} and these magnitudes were converted to flux densities. Using the Rayleigh-Jeans approximation of Planck's radiation law a flux vs. wavelength plot could be produced in order to calibrate the stellar spectra. The average calibration curve obtained independently from both of the standard star observations was used to flux calibrate the science data cube. The telluric contamination present in the observation was also corrected using the standard star observations. Each spectrum present in the science observation was divided by a normalised telluric correction spectrum obtained from the standard star HD 73105.  

\subsection{Continuum subtraction and scattered emission removal} \label{sec: continuum+scattered removal}

\begin{figure*}
    \centering
    \includegraphics[width=18cm, trim= 0cm 0cm 0cm 0cm, clip=true]{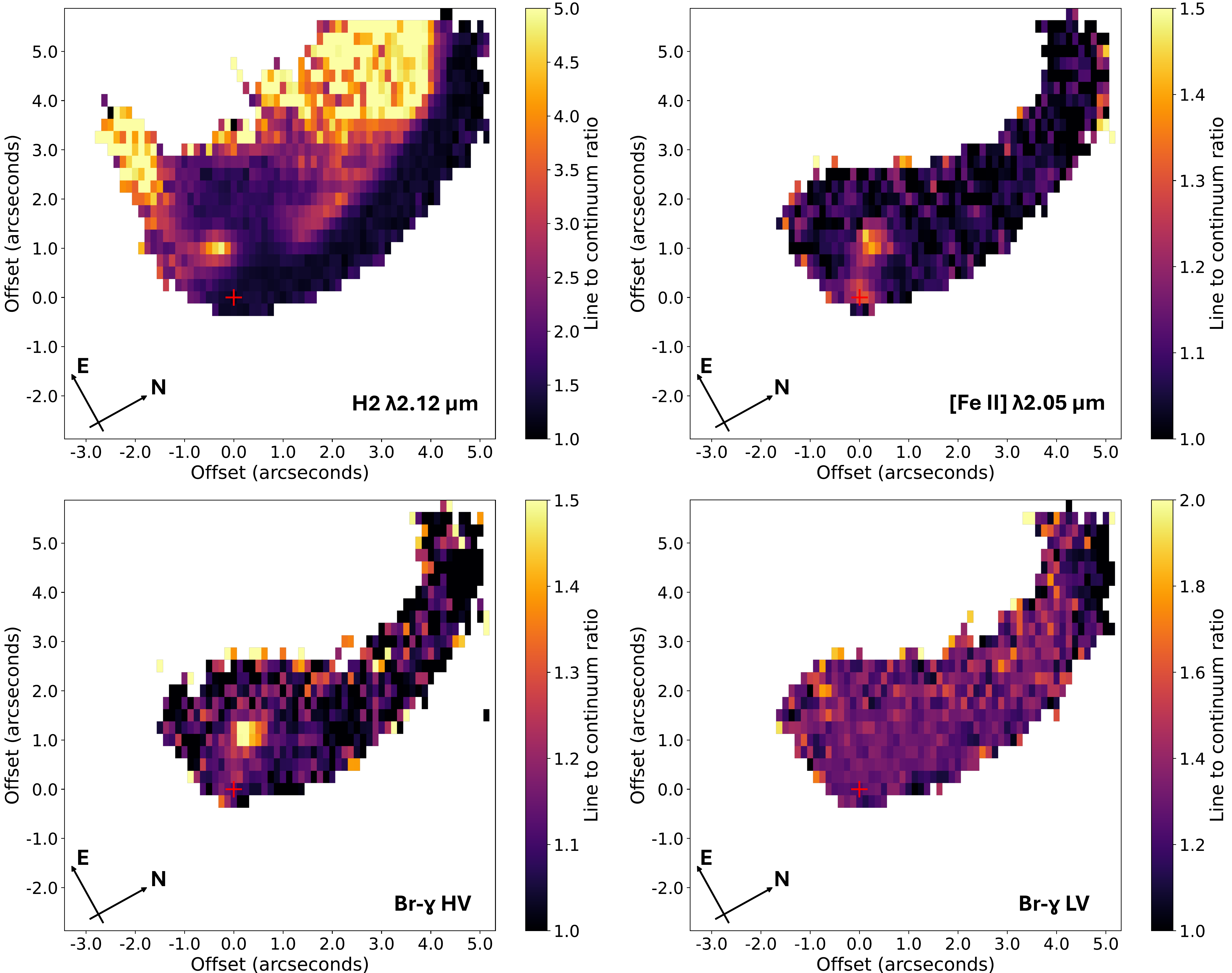}
    \caption{Line-to-continuum ratio plots of H$_2$, [\ion{Fe}{ii}], and Br-$\gamma$ emission. Intrinsic emission appears bright in these images, whereas scattered light is represented by a constant ratio close to unity. The source position is represented with a red plus symbol. A signal-to-noise threshold, set at three times the background noise, is implemented in these images to reject background emission, meaning the morphologies one sees in these figures is due to a combination of the continuum and the intrinsic line emission.
    Top left: Integrated H$_2$ emission between $-$140 km~s$^{-1}$ and +60 km~s$^{-1}$ divided by nearby continuum. The intrinsic line emission here comes from the cavity walls and a H$_2$ knot at the base of the left cavity wall (\textit{A1}).
    Top right: Integrated [\ion{Fe}{ii}] emission between $-$310 km~s$^{-1}$ and $-$100 km~s$^{-1}$ divided by nearby continuum. The intrinsic line emission here traces the high-velocity atomic jet.
    Bottom left: Integrated high-velocity Br-$\gamma$ emission between $-$230 km~s$^{-1}$ and $-$130 km~s$^{-1}$ divided by nearby continuum. The intrinsic line emission here again traces the high-velocity atomic jet. 
    Bottom right: Integrated low-velocity Br-$\gamma$ emission between $-$100 km~s$^{-1}$ and +100 km~s$^{-1}$ divided by nearby continuum. There is no spatially extended intrinsic emission detected here. }
\label{linetocontinuum}
\end{figure*}

The top panel of Fig. \ref{Morphologyfig} in Sect. \ref{HH4647} displays a continuum image, integrated over $\lambda$2.190 $\mu$m - $\lambda$2.191 $\mu$m, which represents the same number of data cube slices as the integrated H$_2$ and [\ion{Fe}{ii}] images displayed in the central and bottom panels. The continuum emission is tracing a reflection nebula with an approximately parabolic shape that is asymmetric in intensity about the average jet PA of 52.3\textdegree\ from \cite{Erkal2021} indicated by a green arrow in the top panel of Fig. \ref{Morphologyfig}. The position of the driving source of the outflow is marked with a red plus symbol. The morphology of the continuum emission delineates the boundary region between the outflow and the ambient material, sweeping out a wide-angled hollow cavity in the quiescent material. The boundary region one sees is essentially the cavity walls. The northern arm of the cavity extends further from the source than the eastern arm, suggesting a dissymmetry in the distribution of ambient material surrounding the outflow. This reflection nebula is light from the driving source of the outflow scattered off the walls of the cavity, revealing the extent of the scattered light component present in these images, this contribution is present at all wavelengths. 

The observations presented here probe the launching region of an outflow, close to the embedded driving source. Strong contributions of scattered light from the central engine is expected in these regions. As we endeavoured to investigate spectro-images of the outflow in intrinsic line emission, the scattered emission first needed to be removed. As the cavity was fully illuminated in continuum emission, an assumption was made that the scattered light component was entirely light from the unresolved central binary reflecting off the cavity. A spectrum was extracted from the continuum peak position (or the source position) to best represent the scattered emission contribution. A polynomial fit was applied to this source spectrum in order to rescale the source spectrum to the same level as the continuum at each spatial-pixel (spaxel) in the data cube. Subsequently, the rescaled source spectrum was subtracted locally in wavelength around each emission line of interest from the spectra at each each spaxel. This effectively removed the scattered light component and continuum emission from the emission line of interest, leaving only the intrinsic line emission. This scattered emission removal routine was performed locally in the spectral region around each emission line of interest to ensure an accurate subtraction of scattered light. 

Line to continuum ratio plots are an effective tool for investigating the contribution of scattered emission and for checking the effectiveness of any process for removing it. These plots display the result of calculating the ratio between an integrated line image and an integrated continuum image, taken spectrally nearby to the line of interest. The contribution from scattered light to the integrated line images will be represented by a constant ratio in these plots, as the scattered light follows the same spatial distribution as the continuum emission. In contrast, the real, intrinsic local emission will appear bright in these images, exceeding the constant ratio that represents the contribution from scattered light. The top left panel of Fig. \ref{linetocontinuum} displays the result for H$_2$ emission. It displays ratio values approximately between 1 and 2 in the inner region of the cavity, close to the source, and along the outer side of the northern pointing cavity wall. Here, the contribution from scattered light is strongest. It is seen that the intrinsic line emission here is coming from the cavity walls and the H$_2$ knots at the bases of the cavity walls, in agreement with the central panel of Fig. \ref{Morphologyfig}. Intrinsic emission appears brighter as offset from the source increases as the contribution from scattered light decreases with distance from the source. This is why the tips of the northern and eastern cavity walls have significantly larger ratio values. Figure \ref{Scattered_emission_comparison} displays the integrated intensity map of the H$_2$ emission before and after the scattered emission removal routine was implemented. When comparing between these two images, it is clear that after the scattered light is subtracted, there is a steeper intensity gradient associated with the intrinsic H$_2$ emission, which is most readily observable in the closeness of contour lines surrounding bright H$_2$ emission, such as the features labelled \textit{A1} \& \textit{A2}. Nebulous emission close the apex of the cavity has been diminished, which is in agreement with H$_2$ line to continuum map predicting a strong contribution from scattered photons in this region.

As the scattered emission subtraction process samples the representative scattered emission spectrum from the continuum peak, the base of the jet, traced in forbidden iron emission and Br-$\gamma$ emission, is removed as a unavoidable by-product of the process, as jet emission is more compact and closer to the source position. This implies that some intrinsic atomic jet emission is present in the representative scattered emission spectrum or source spectrum due to the diameter of the aperture used to sample the spectrum, and also the angular resolution of the observation. 

The top right panel of Fig. \ref{linetocontinuum} presents the line to continuum plot for the [\ion{Fe}{ii}] emission. At the continuum peak position one sees that the line to continuum ratio is distinctly higher than that in the surrounding nebulous region, confirming that there is intrinsic jet emission co-located with the continuum peak. Therefore, we cannot use the scattered emission spectrum extracted from the continuum peak position to subtract the scattered light in [\ion{Fe}{ii}] emission as we would overcorrect and subtract intrinsic jet emission. This is further supported in Fig. 6 from \cite{Nisini2024}. The [\ion{Fe}{ii}] $\lambda$1.87~$\mu$m jet as observed using NIRCam on JWST is displayed in the top right panel. The first jet knot one sees is offset from the binary source, separated only by a thin dark lane. The angular resolution provided by the SINFONI observations presented here do not spatially resolve the [\ion{Fe}{ii}] jet knot from the central binary. This results in an artificially increased line to continuum ratio at the source position, as jet emission coincides with the source position, where light from the source alone is contributing to scattered light in the nebulous cavity. In contrast, the bottom left panel of Fig. 6 from \cite{Nisini2024} highlights that the innermost bright peak in H$_2$ emission, \textit{A1}, is sufficiently separated from the binary source to be spatially resolved by SINFONI. This justifies that the spectrum extracted from the source position is representative of the true H$_2$ stellar spectrum illuminating the cavity, and it can therefore be used for correcting the scattered light in H$_2$ emission. As we were sacrificing the ability to trace the jet back to the source when applying the scattered emission removal process to the jet tracing emission lines, we chose to not perform this routine on the jet tracing lines. Furthermore, the lack of scattered emission removal is not critical in our analysis of the jet tracing lines because the narrow and collimated jet emission stands out clearly from the fainter surrounding nebulous emission allowing us to accurately determine the jet PA and morphology.

The bottom left panel of Fig. \ref{linetocontinuum} presents the high-velocity ($-$230 km~s$^{-1}$ to $-$130 km~s$^{-1}$) Br-$\gamma$ emission line to continuum plot. This emission is seen to trace the jet also. Therefore, scattered emission subtraction was not performed on this emission line either. A baseline continuum subtraction routine was instead implemented by performing local polynomial fitting to the continuum around the emission lines. This polynomial fit continuum subtraction was performed at every spaxel in the data cube, for each emission line, with care taken to exclude the line emission region from the baseline fit. 

In the bottom right panel of Fig. \ref{linetocontinuum}, we present a line to continuum map of  low-velocity Br-$\gamma$ emission, integrated from $-$100 km~s$^{-1}$ to +100 km~s$^{-1}$. In this line to continuum map, the ratio of integrated line emission to continuum is approximately constant across the morphology of the cavity, and equal to the ratio located at the continuum peak position. We can infer from this that any spatially extended Br-$\gamma$ emission one sees in the low-velocity interval follows the same spatial distribution as the scattered light seen in continuum. This implies that any extended low-velocity Br-$\gamma$ emission is not intrinsic emission and is instead dominated by scattering of the central source spectrum off of the dusty, nebulous cavity material traced by the continuum. 

The work presented in this paper highlights that for similar studies of outflows from embedded sources, it is critical to consider the contribution of scattered light to the emission line profiles, and to acknowledge that removing the continuum emission using a baseline subtraction could strongly overestimate the true extent of in-situ emission in some lines. Examples here are the H$_2$ emission and the low-velocity Br-$\gamma$ emission.

\section{Results} \label{sec: results}

\begin{figure*}
    \centering
    \includegraphics[width=16cm, trim= 0cm 0cm 0cm 0cm, clip=true]{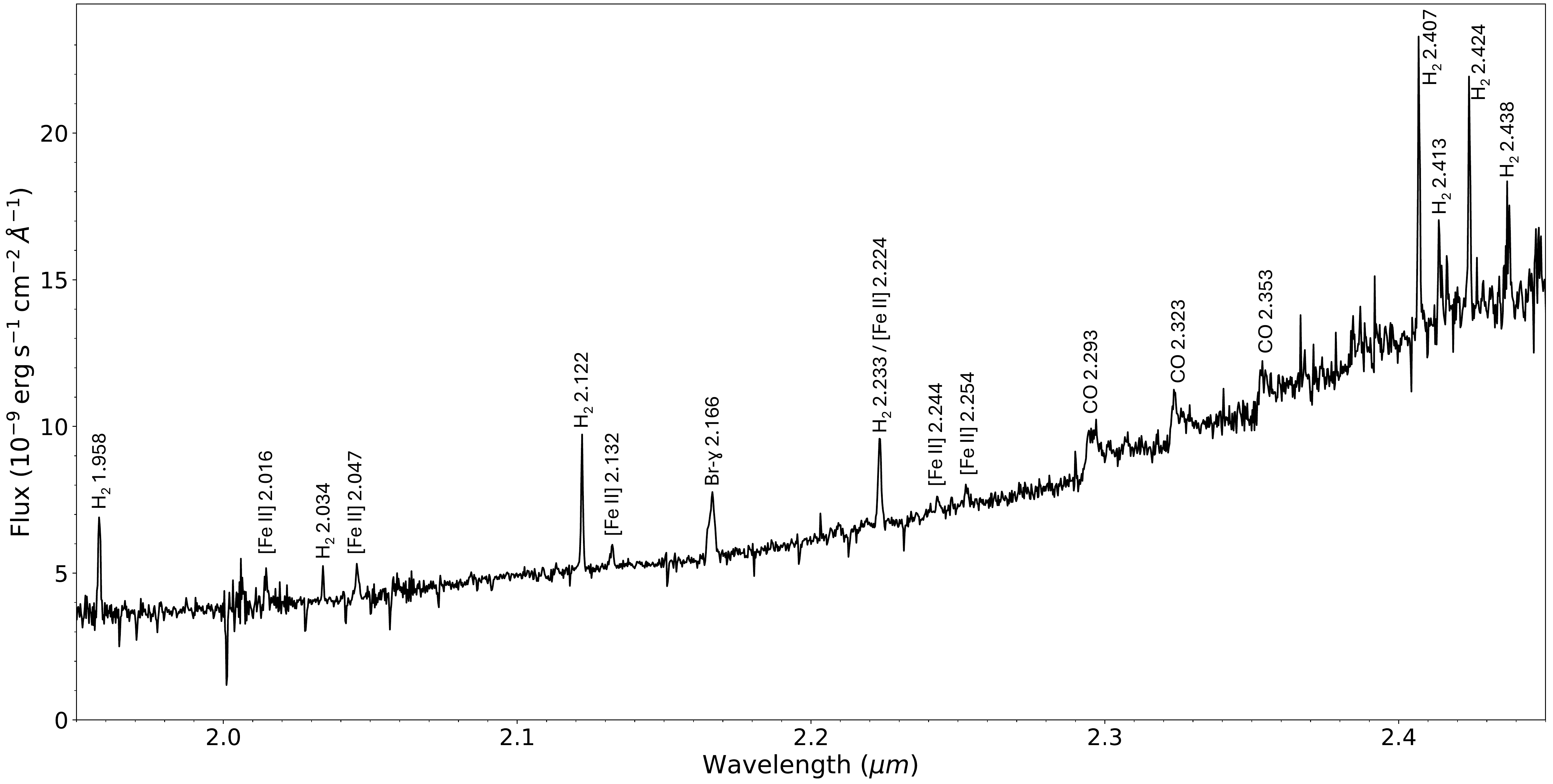}
    \caption{Spectrum extracted using a circular aperture of radius 0\farcs41 from the continuum peak. The jet emission is identified in the FELs of [\ion{Fe}{ii}], and in Br-$\gamma$ emission. The outflow is detected in the H$_2$ emission lines. In addition, CO overtone bandhead emission at 2.29-2.36~$\mu$m, which traces the inner gaseous disc, is also seen. A characteristic reddening slope is evident in this spectrum as expected, as the source is heavily extincted and only visible at infrared wavelengths. Stellar absorption lines were not identified in this spectrum, it was concluded that the central source is too heavily extincted to study these lines.}
    \label{Stellarspec}
\end{figure*}

The SINFONI observation of HH 46/47 reveals emission lines tracing outflow and accretion activity, in agreement with previous spectroscopic studies \citep{Antoniucci2008, Garcia-Lopez2010}. In this paper we discuss the morphology and kinematics of the jet and wide-angled outflow with a particular focus on the kinematics. Figure \ref{Morphologyfig} presents integrated intensity maps that showcase the morphology of the HH 46/47 blueshifted outflow. Maps of the continuum emission, molecular hydrogen (H$_2$) emission at 2.12 $\mu$m, and forbidden iron emission at [\ion{Fe}{ii}] $\lambda$2.047~$\mu$m are shown. The H$_2$ emission line at 2.12 $\mu$m, hereafter referred to as H$_2$, is the brightest MHEL in the dataset that does not suffer any negative effects due to closeness to the spectral edge of the detector. The forbidden iron emission line at 2.047 $\mu$m, hereafter referred to as [\ion{Fe}{ii}], is the brightest iron emission line in the dataset. The source is partially embedded at these wavelengths but we detect a point source at the apex of the cavity in continuum emission. We fit a 2D Gaussian profile to this continuum peak and use this as our driving source position throughout. This continuum peak is not necessarily coincident with the position of the HH 46 IRS millimetre position derived by \cite{Arce2013}, as in the IR it is possible that the significant contribution of scattered light could shift the barycentre of the source continuum peak. A spectrum extracted from the source position is presented in Fig. \ref{Stellarspec}. Atomic jet emission is identified in the forbidden emission lines (FELs) of [\ion{Fe}{ii}], and in Br-$\gamma$ emission. Several H$_2$ ro-vibrational emission lines ($v$ = 1-0 \& 2-1) are detected, likely due to a wide-angled outflow cavity (see Sect. \ref{sec:morphology}). In addition, CO overtone bandhead emission at 2.29 - 2.36~$\mu$m, which traces the inner gaseous disc, is also seen.\footnote{Spectra extracted from an atomic jet knot at 1\farcs0 (450 au) offset from the source position, and from both emission peaks (\textit{A1} \& \textit{A2}) either side of the V-shaped cavity traced by molecular hydrogen (H$_2$) are also presented in Fig. \ref{HH4647spectra}.}

\subsection{Morphology} \label{sec:morphology}

The morphology of the integrated continuum image displayed in the top panel of Fig. \ref{Morphologyfig} was discussed in Sect. \ref{sec: continuum+scattered removal}. The structure of the cavity as traced by continuum emission, H$_2$ emission, and CO emission is further compared in Sect. \ref{sec: layeringOfEmission}. The central panel of Fig. \ref{Morphologyfig} displays the continuum subtracted and scattered emission subtracted H$_2$ 1-0 S(1) emission, integrated over radial velocities of $v$ = $-$95 km~s$^{-1}$ to +110 km~s$^{-1}$. As also shown by the JWST NIRSpec H$_2$ 2.12~$\mu$m image in \cite{Nisini2024}, the integrated H$_2$ emission has a striking V-shaped morphology tracing the limb-brightened cavity walls of the outflow. The shocked northern cavity wall traced by H$_2$ emission lies within the cavity structure seen in the continuum image, meaning the H$_2$ emission is confined within the outflow cavity seen in scattered light. This narrowing of the cavity walls suggests that the H$_2$ emission is probing deeper into the outflow cavity, allowing one to see within the cavity walls traced in scattered light. The H$_2$ cavity is again asymmetric about the jet PA with the northern wall appearing significantly brighter and extending further than the eastern wall. There is also some faint emission present within the cavity, especially apparent close to the interior of northern cavity wall. This emission within the cavity walls signifies that the cavity structure one sees has a 3D inverted-conical shape extending from the source and surrounding the atomic jet. The jet is not detected in H$_2$. A bright H$_2$ knot of emission is also seen at the base of the eastern cavity wall at a distance of approximately 1\farcs0 (450 au) from the source position. We label this bright H$_2$ peak \textit{A1} following \cite{Nisini2024}. We also label the dimmer H$_2$ peak at the base of the northern cavity wall at approximately 1\farcs4 from the source position \textit{A2}, and the arc-like structure extending from the northern cavity wall, curling towards the jet axis \textit{A3}, again, following \cite{Nisini2024}.

The bottom panel of Fig. \ref{Morphologyfig} is the continuum subtracted, but not scattered emission subtracted, [\ion{Fe}{ii}] emission at 2.047 $\mu$m, integrated over radial velocities of $v$ = $-$280 km~s$^{-1}$ to $-$65 km~s$^{-1}$. [\ion{Fe}{ii}] forbidden emission lines are well-established atomic jet tracers. The [\ion{Fe}{ii}] emission traces the base of the high-velocity collimated jet, culminating in a bright knot or bow shock approximately 1\farcs0 (450 au) from the source position. The jet bends slightly, tending towards the northern cavity wall, as the jet PA (52.3\textdegree) and CO outflow PA (60\textdegree) are different from one another. An interesting observation is that the jet knot we trace with [\ion{Fe}{ii}] does not coincide with the bright knot, \textit{A1}, seen in H$_2$ emission, as seen in \cite{Nisini2024}. There is a horizontal offset of approximately 0\farcs53 (240 au) between the centres of these outflow emission knots. This offset between the jet knot seen in [\ion{Fe}{ii}] and the outflow knot, \textit{A1}, seen in H$_2$ suggests that the atomic jet emanates within the H$_2$ cavity, extending between the bases of the cavity walls seen in H$_2$. Again, this indicates that the conical outflow cavity traced by molecular hydrogen surrounds the more collimated and faster moving atomic jet. The morphology of the high-velocity ($-$230 km~s$^{-1}$ to $-$130 km~s$^{-1}$) Br-$\gamma$ emission is almost identical to that traced by [\ion{Fe}{ii}], displaying the base of the jet and the same knot at 1\farcs0 from the source position. This is shown in the bottom left panel of Fig. \ref{linetocontinuum}. The Br-$\gamma$ emission is discussed further in Sect. \ref{kinematics}. 

%Br-$\gamma$ emission is overplotted with [\ion{Fe}{ii}]$\lambda$2.047$\mu$m in figure \ref{velgrad} (right).

\subsection{Kinematics} \label{kinematics}
\begin{figure}[hbt!]
   \centering
    \includegraphics[width=8.3cm, trim= 0cm 0cm 0cm 0cm, clip=true]{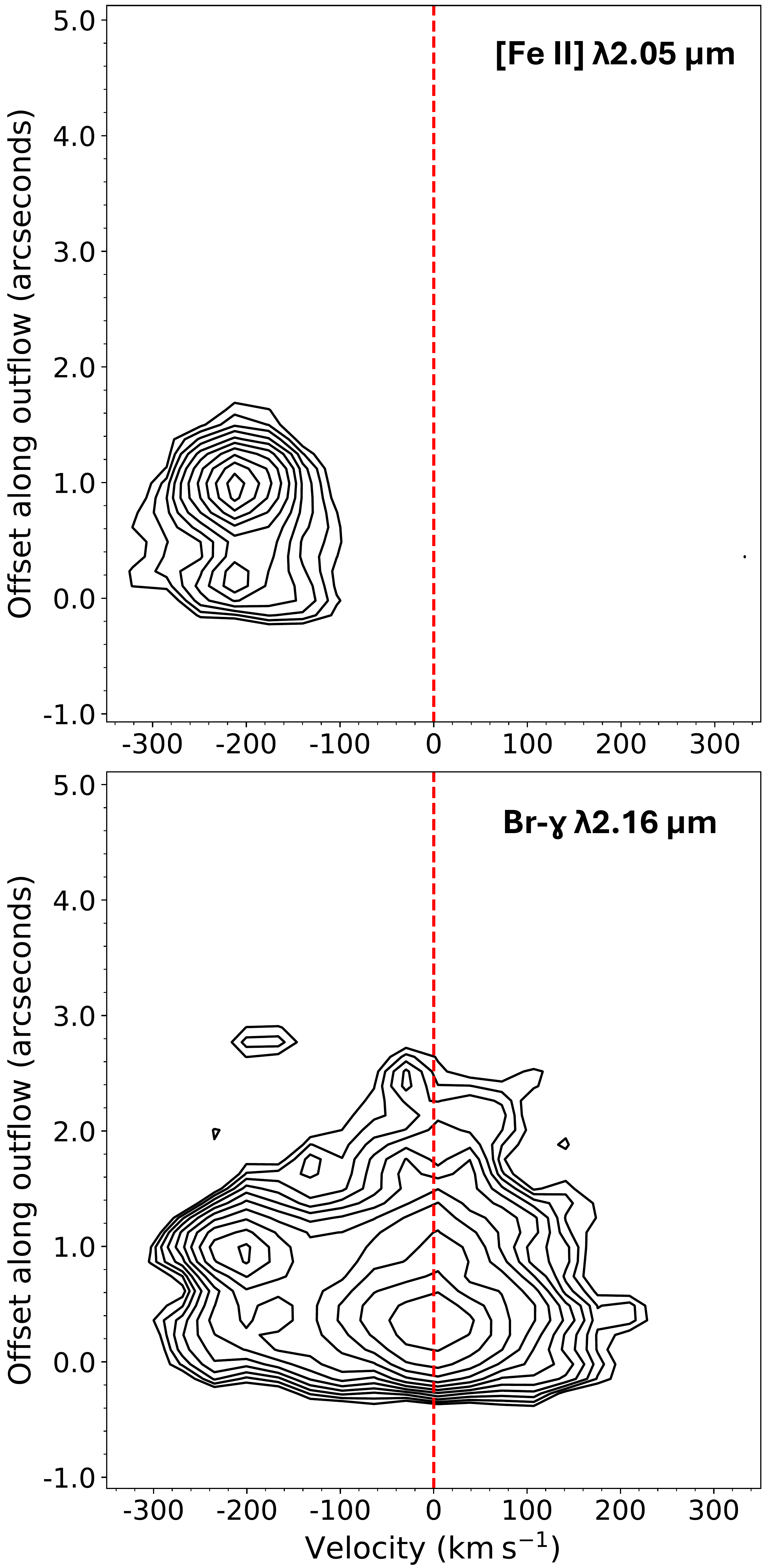}
    \caption{Position velocity diagrams extracted along the outflow axis (PA = 60 \textdegree) with a pseudo slit width of 1\arcsec. Upper panel: Position velocity diagram of the [\ion{Fe}{ii}] emission line region. The emission is traced from the source at 0\arcsec~and culminates in a bow shock at $\sim$ 1\farcs0 (450~au). The radial velocity of the jet emission traced in this line peaks at approximately $-$200 km~s$^{-1}$, in agreement with previous observations. Lower panel: Position velocity diagram of the Brackett-$\gamma$ emission line region. The emission is again traced from the source and culminates in a bow shock at $\sim$ 1\farcs0 (450~au). The radial velocity of the jet emission traced in this line peaks at approximately $-$200 km~s$^{-1}$. The spatially extended low-velocity Br-$\gamma$ emission here, that is, the emission centred around 0 km~s$^{-1}$ and having a full width extending to approximately $\pm$ 180 km~s$^{-1}$, is not intrinsic emission, it is instead light scattered off of the nebulous cavity material. Contour levels for both begin at 3$\sigma$ of the background emission and increase by factors of 1.2. Velocities are expressed with respect to the LSR.}
    \label{pv}
\end{figure}

Position velocity (PV) diagrams can be extracted from integral-field spectroscopic data. A pseudo-slit is placed along the desired emission region, the spatial information within the slit is collapsed and represented by one axis and the wavelength information is converted into a radial velocity using Doppler shift and represented on the other axis. Figure \ref{pv} displays two PV diagrams, taken along the outflow axis, with a PA of 60\textdegree, from \cite{Zhang2019}, and a pseudo-slit width of 1\arcsec.

\begin{figure}
   \centering
    \includegraphics[width=8.8cm, trim= 0cm 0cm 0cm 0cm, clip=true]{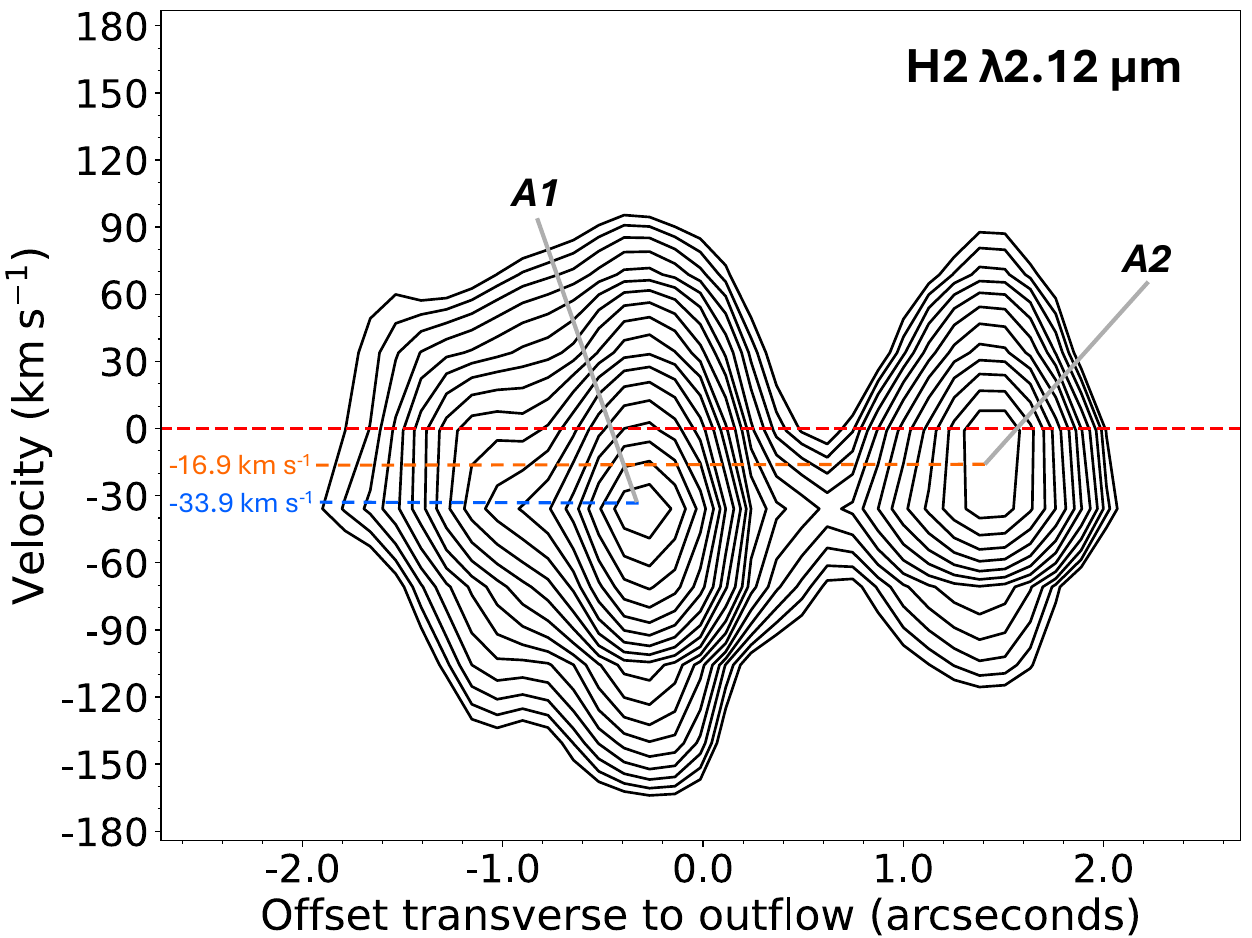}
    \caption{Transverse PV diagram of the H$_2$ emission line region extracted at a projected \textit{z} of 1\farcs2 and with a pseudo slit width of 0\farcs5. The radial velocity of the molecular emission is much lower than that of the atomic emission. Two emission peaks can be seen on either side of the source position, labelled \textit{A1} and \textit{A2}. The emission peak on the left side of the cavity (\textit{A1}) has a slightly higher radial velocity of $-$33.9 km~s$^{-1}$ than the emission peak on the right side of the cavity (\textit{A2}) at $-$16.9 km~s$^{-1}$. This hints at the presence of a velocity gradient transverse to the outflow direction in the cavity. Contour levels begin at 3$\sigma$ of the background emission and increase by factors of 1.2.}
    \label{pvtransverse}
\end{figure}

The top image in Fig. \ref{pv} displays a PV diagram extracted along the PA of the CO outflow axis in [\ion{Fe}{ii}] emission. Here one sees emission traced from the source position at 0\arcsec~culminating in a knot at $\sim$ 1\farcs0 or 450~au. The radial velocity of the emission peaks at approximately $-$200 km~s$^{-1}$, as is expected for a collimated atomic jet. A Gaussian profile was fit to the emission peak at offset 1\farcs0, along the velocity axis, and a Gaussian centroid velocity of $v_{\mathrm{LSR}}$ = $-$208.7 $\pm$ 3.6 km~s$^{-1}$ was retrieved. These radial velocities are consistent with previous observations \citep{Garcia-Lopez2010, Nisini2024}. In the bottom image of Fig. \ref{pv} a PV diagram, again taken along the PA of the CO outflow axis is shown, traced by Brackett-$\gamma$. The high-velocity emission ($\sim$ $-$200 km~s$^{-1}$) in this PV diagram is almost identical to that traced by [\ion{Fe}{ii}] implying that they are both tracing the jet. The high-velocity emission again culminates in a knot at $\sim$ 1\farcs0 or 450~au. Again, a Gaussian profile was fit to the emission peak at offset 1\farcs0, along the velocity axis, and a Gaussian centroid velocity of $v_{\mathrm{LSR}}$ = $-$211.9 $\pm$ 3.5 km~s$^{-1}$ was retrieved. A strong low-velocity component at $\sim$ 0~km~s$^{-1}$ with a full width extending to approximately $\pm$ 180 km~s$^{-1}$ is also evident. As discussed in Sect. \ref{sec: continuum+scattered removal} the line to continuum map of this low-velocity Br-$\gamma$ emission, displayed in Fig. \ref{linetocontinuum}, implies that all of the extended low-velocity Br-$\gamma$ emission is not intrinsic emission and is instead dominated by scattering. The intrinsic emission traced at approximately 0 km~s$^{-1}$ and having a full width extending to approximately $\pm$ 180 km~s$^{-1}$, is very compact, spatially unresolved emission entirely collocated at the source position, possibly tracing accretion processes and compact disc winds.

In Fig. \ref{pvtransverse} a PV diagram taken transverse to the outflow direction at a PA of 150\textdegree\ and at a distance of 1\farcs2 above the source position is shown, traced by H$_2$ emission after scattered emission removal. This PA is perpendicular to the PA of the CO outflow axis. The width of the pseudo-slit used to extract this PV diagram is 0\farcs5. The pseudo-slits position on the H$_2$ cavity is shown in the bottom panel of Fig. \ref{Scattered_emission_comparison}. Notice that the radial velocity of material traced by H$_2$ is much lower than that traced by [\ion{Fe}{ii}] emission and Br-$\gamma$. Two emission peaks are seen either side of the source position, labelled \textit{A1} and \textit{A2}. These emission peaks are the bases of each outflow cavity wall seen in H$_2$ emission in Fig. \ref{Morphologyfig}. One sees that the emission peak from the eastern cavity wall (\textit{A1}) peaks at a slightly higher radial velocity than the emission from the northern cavity wall (\textit{A2}). A Gaussian fit was applied to both emission peaks in the velocity axis to retrieve the Gaussian centre of the emission. The Gaussian centroid velocity ($v_{\mathrm{LSR}}$) of the emission peak \textit{A1} is $-$33.9 $\pm$ 1.4 km~s$^{-1}$. The Gaussian centroid velocity of the emission peak \textit{A2} is $-$16.9 $\pm$ 1.5 km~s$^{-1}$. These are highlighted on the figure. This difference in centroid velocity between the two sides of the cavity walls hints at the presence of a velocity gradient in the molecular cavity, transverse to the outflow direction.

\subsection{H$_2$ velocity centroid map} \label{Sec: H2 centroid map}

\begin{figure*}
\centering
    \includegraphics[width=18cm, trim= 0cm 0cm 0cm 0cm, clip=true]{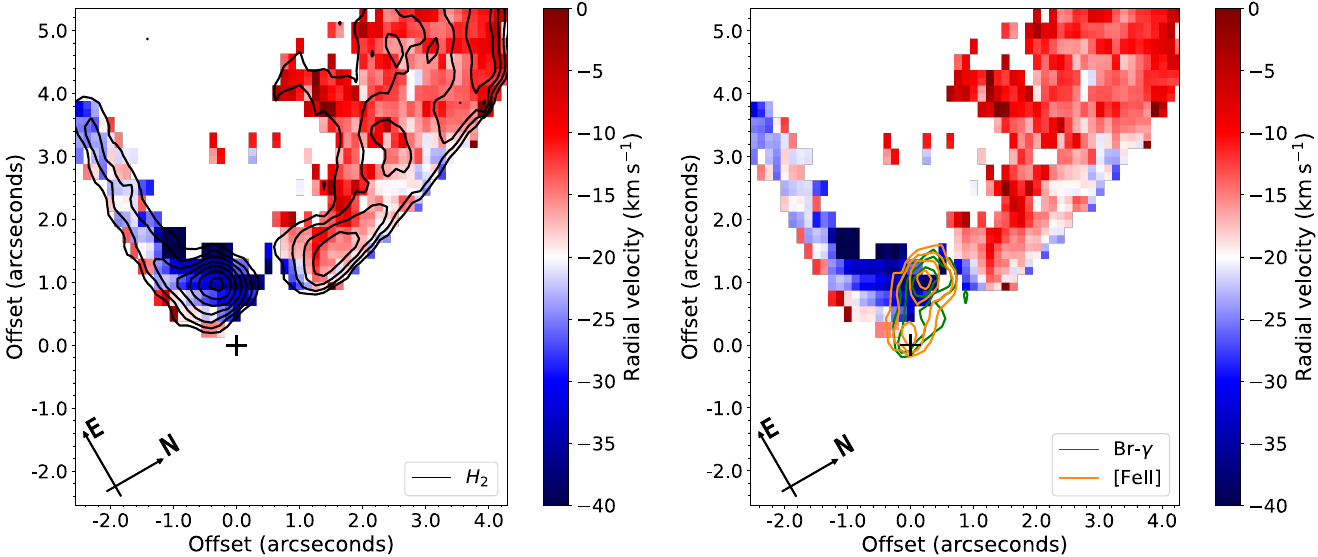}
    \caption{Scattered light-corrected H$_2$ emission radial velocity centroid maps. The source position is marked by a black plus symbol. The radial velocity centroid maps display a clear velocity gradient between both sides of the V-shaped cavity. The left side of the cavity has, on average, a greater radial velocity when compared with the right side. The difference in velocity between both sides is $\sim$ 10 km~s$^{-1}$. Contour levels for both begin at 3$\sigma$ of the background emission and increase by factors of 1.5.
    Left: H$_2$ velocity centroid map with integrated H$_2$ emission overplotted in black contour lines. Right: Same H$_2$ velocity centroid map but with overlayed plots of jet-tracing integrated [\ion{Fe}{ii}] emission (in orange contours) and integrated Br-$\gamma$ emission (in green contours). The jet flows between both sides of the cavity traced by H$_2$.}
    \label{velgrad}
\end{figure*}

The transverse PV diagram in H$_2$ emission hinting that both sides of the V-shaped outflow cavity had different radial velocities prompted the production of a 2D radial velocity map of the H$_2$ emission after scattered emission removal. The H$_2$ emission line at 2.122 $\mu$m is chosen here as it possesses the highest S/N of all of the H$_2$ emission lines in our observation and does not suffer any erroneous effects that are common near the spectral edges of IFU observations. The production of a 2D radial velocity map of the H$_2$ emission was achieved by fitting Gaussian profiles to each H$_2$ emission line at every spaxel in the data cube and rejecting fits with a S/N below a certain threshold. The centres of each of the Gaussian profiles were taken as the centroid radial velocities of the emission line. This resulted in a 2-Dimension velocity centroid map of the H$_2$ 2.12~$\mu$m emission region shown in Fig. \ref{velgrad}. The velocities in this map are expressed with respect to the LSR velocity of the cloud that the outflow emanates from. The error associated with the velocity centroid map is a combination of the uncertainty associated with the Gaussian fitting, namely the accuracy of the Gaussian centroid measurement and the uncertainty involved with the OH emission line wavelength recalibration. Thus, summing the relative and absolute errors in quadrature, we derive a 1$\sigma$ uncertainty associated with the velocity centroid map of $\pm$ 1.0 km~s$^{-1}$ to $\pm$ 3.0 km~s$^{-1}$ depending on the brightness of the H$_2$ line.

% We also estimate a 1σ error on the global absolute velocity calibration scale of 10 km s−1. When performing Gaussian fitting to derive the line velocity centroids, errors in centroid velocity are a combination of the error in velocity calibration and that of Gaussian fitting. The former remains constant, while the latter is signal-to-noise dependent. Fitting errors were calculated in accordance with the equation for the accuracy of a Gaussian centroid fit as applied to cases of high signal-to-noise such that photon noise dominates, and assuming that the profile is well-sampled (Whelan & Garcia 2008): σcentroid = FWHM 2 √ 2 ln 2 SNR (1) where FWHM represents the width of the fitted profile, (here our spectral resolution of 56 km s−1), and SNR is the signal-to-noise ratio at the Gaussian peak. For our observations, we thus derive typical centroid velocity 3σ accuracy of 2–5 km s−1, depending on the signal to noise. Combining this in quadrature with the internal relative calibration error gives the error bars

One sees the V-shaped cavity morphology reproduced in velocity centroid maps. The left image in the figure displays the integrated H$_2$ emission intensity map overplotted in black contour lines. The source position is marked with a black plus symbol. Here, $-$20 km~s$^{-1}$ was chosen as the centre point for the radial velocity gradient map. One sees that the left outflow cavity wall has, on average, higher radial velocity peaks when compared to the right wall of the cavity. The velocity difference between both sides of the cavity is $\sim$ 10 km~s$^{-1}$. In order to recover a more quantitative value for the radial velocity difference between both sides of the H$_2$ cavity, transverse PV cuts, similar to that presented in Fig \ref{pvtransverse}, were taken at different heights ($z$) along the outflow. PV cuts of width 0\farcs25 were taken at $z$ = 1\farcs0, 1\farcs5, 2\farcs0, 2\farcs5, and 3\farcs0. The emission peaks corresponding to each side of the H$_2$ cavity in each PV plot were fit with Gaussian profiles along the velocity axis to retrieve the centroid velocity in the same way the Gaussian centroid velocities were retrieved in Fig. \ref{pvtransverse}. The difference in radial velocity centroid measurements ($\Delta v_{\mathrm{rad}}$) between the left side of the H$_2$ cavity ($v_{\mathrm{rad,\ left}}$) and the right side ($v_{\mathrm{rad,\ right}}$), at each height ($z$) along the outflow, is presented in Table \ref{Table:Velocity difference}. The uncertainties associated with the centroid velocities are quadrature summations of the Gaussian centroid error and the uncertainty obtained with wavelength recalibration, as discussed in Sect. \ref{WavelengthRecal}. 

\begin{table}
\caption{Velocity difference between eastern and northern pointing H$_2$ cavity walls at various heights along the outflow.} 
\label{Table:Velocity difference}
    \centering
    \begin{tabular}{cccc}
    \hline\hline  
        $z$ (\arcsec) & $v_{\mathrm{rad,\ left}}$ (km~s$^{-1}$) & $v_{\mathrm{rad,\ right}}$ (km~s$^{-1}$) &$\Delta v_{\mathrm{rad}}$ (km~s$^{-1}$) \\ \hline
        1.0 & 31.3 $\pm$ 1.5 & 17.3 $\pm$ 2.5 & 14.0 $\pm$ 2.9\\ %\cmidrule(l r){1-4}
        1.5 & 25.1 $\pm$ 2.0 & 13.4 $\pm$ 1.4 & 11.7 $\pm$ 2.4\\  %\cmidrule(l r){1-4}
        2.0 & 24.0 $\pm$ 2.1 & 15.2 $\pm$ 1.4 & 8.8 $\pm$ 2.5 \\  %\cmidrule(l r){1-4}
        2.5 & 22.0 $\pm$ 2.3 & 16.6 $\pm$ 2.9 & 5.4 $\pm$ 3.7 \\  %\cmidrule(l r){1-4}
        3.0 & 22.8 $\pm$ 2.5 & 14.5 $\pm$ 2.0 & 8.3 $\pm$ 3.2 \\  %\cmidrule(l r){1-4}
        \hline
    \end{tabular}
    %\captionsetup{justification=centering}
    %\captionsetup{tableposition=top}
\end{table}

The right image in Fig. \ref{velgrad} presents the same H$_2$ velocity centroid map, but integrated jet emission traced by [\ion{Fe}{ii}] in orange contours, and integrated Br-$\gamma$ emission in green contours are overplotted instead. These emission lines are tracing the base of the jet. One sees that the jet emission flows in-between either side of the H$_2$ cavity walls. This further illustrates that the H$_2$ low-velocity gas is surrounding the high-velocity jet in a conical morphology. This velocity gradient pattern observed between the eastern and northern pointing walls of the outflow cavity is also seen in the other molecular hydrogen emission lines present in our observation.

% \begin{figure*}
%    \centering
%     \includegraphics[width=18cm, trim= 0cm 0cm 0cm 0cm, clip=true]{images/H2velgradfigure.pdf}
%     \caption{H$_2$ $\lambda$2.122$\mu$m radial velocity centroid maps. The source position is marked by a black cross. The radial velocity centroid maps display a clear velocity gradient between both sides of V-shaped cavity. The left side of the cavity has, on average, a greater radial velocity when compared with the right side. The difference in velocity between both sides is $\sim$ 10 km~s$^{-1}$. Contour levels for both begin at 3$\sigma$ of the background emission and increase by factors of 1.5.
%     \textit{Left:} H$_2$ $\lambda$2.122$\mu$m velocity centroid map with integrated H$_2$ $\lambda$2.122$\mu$m emission overplotted in black contour lines.  \textit{Right:} The same H$_2$ $\lambda$2.122$\mu$m velocity centroid map with jet-tracing integrated [\ion{Fe}{ii}]$\lambda$2.047$\mu$m emission in orange contours, and integrated Br-$\gamma$ emission in green contours overplotted. The jet flows between both sides of the cavity traced by H$_2$ implying that the H$_2$ emission surrounds the jet in a cone-like morphology. The base of the jet is also deflected away from the left side of the cavity.}

% \end{figure*}   

% \begin{figure}
%    \centering
%   \includegraphics[width=10cm, trim= 0cm 0cm 0cm 0cm, clip=true]{images/H2velocitygrad_H2.pdf}
%     \caption{\textbf{BEFORE UNEVEN SLIT CORRECTION} H$_2$ $\lambda$2.122$\mu$m velocity centroid map displaying clear velocity gradient between both sides of V-shaped cavity. Integrated H$_2$ $\lambda$2.122$\mu$m emission intensity map is overplotted in black contour lines and the source position is marked with a black cross. The left side of the cavity has, on average, a greater radial velocity when compared with the right side. The difference in velocity between both sides is $\sim$ 10 km~s$^{-1}$.}
%     \label{H2velgrad_H2}
%     \end{figure}

% \begin{figure}
%    \centering
%   \includegraphics[width=10cm, trim= 0cm 0cm 0cm 0cm, clip=true]{images/H2velocitygrad_jet.pdf}
%     \caption{\textbf{BEFORE UNEVEN SLIT CORRECTION} H$_2$ $\lambda$2.122$\mu$m velocity centroid map. Jet-tracing integrated forbidden iron emission ([\ion{Fe}{ii}]$\lambda$2.047$\mu$m) in orange contours, and integrated Br-$\gamma$ emission in green contours are overplotted. The source position is marked with a black cross. The jet bisects the cavity, flowing between both sides implying that the H$_2$ emission surrounds the jet. The base of the jet is also deflected away from the left side of the cavity.} 
%     \label{H2velgrad_jet}
%     \end{figure}

\section{Discussion} \label{sec: Discussion}
\cite{Nisini2024} present spectacular JWST NIRSpec and MIRI observations of the HH 46/47 molecular outflow, cavity and atomic jet. Their NIRSpec 6\arcsec\ $\times$ 6\arcsec\ FOV encompasses a similar region to the SINFONI observations presented here but they have the advantage of the superior angular resolution of JWST. They suggest that \textit{A1} is a expanding bubble ejected by the secondary component of the binary system, as the \textit{A1} knot has a shell-like morphology revealed by the higher NIRSpec resolution and additional NIRCam images. They associate both \textit{A2} and \textit{A3} with material entrained by the primary jet. The morphology of the H$_{2}$ emission obtained with SINFONI and seen in Fig. \ref{Morphologyfig} further supports these conclusions. However, we cannot rule out with the SINFONI data alone that \textit{A1} and \textit{A2} instead trace the base of the cavity walls surrounding the jet. The spectral resolution of the \cite{Nisini2024} observations is R $\sim$ 2700, which translates to $\sim$ 110 km~s$^{-1}$ of velocity resolution. The spectral resolution provided with SINFONI in the K-band (R $\sim$ 4000) is superior, and the significant advantage of being able to perform a wavelength recalibration with OH lines (as described in Sect. \ref{WavelengthRecal}) means that we can extract important kinematical information about the molecular emission and the atomic jet, which is not possible from the JWST observations.

\subsection{Layering of emission and axial asymmetries} \label{sec: layeringOfEmission}

% {\bf This section is a bit confusing as there is a lot of repetition of information about the jet and cavity walls both within the section and with information given earlier in the manuscript. Also it is a bit confusing to start the discussion of Figure 5 showing the HST data with the inset - I would start by describing the figure. - I have tried to re do the text }

\begin{figure*}
   \centering
   \includegraphics[width=16cm, trim= 0cm 0cm 0cm 0cm, clip=true]{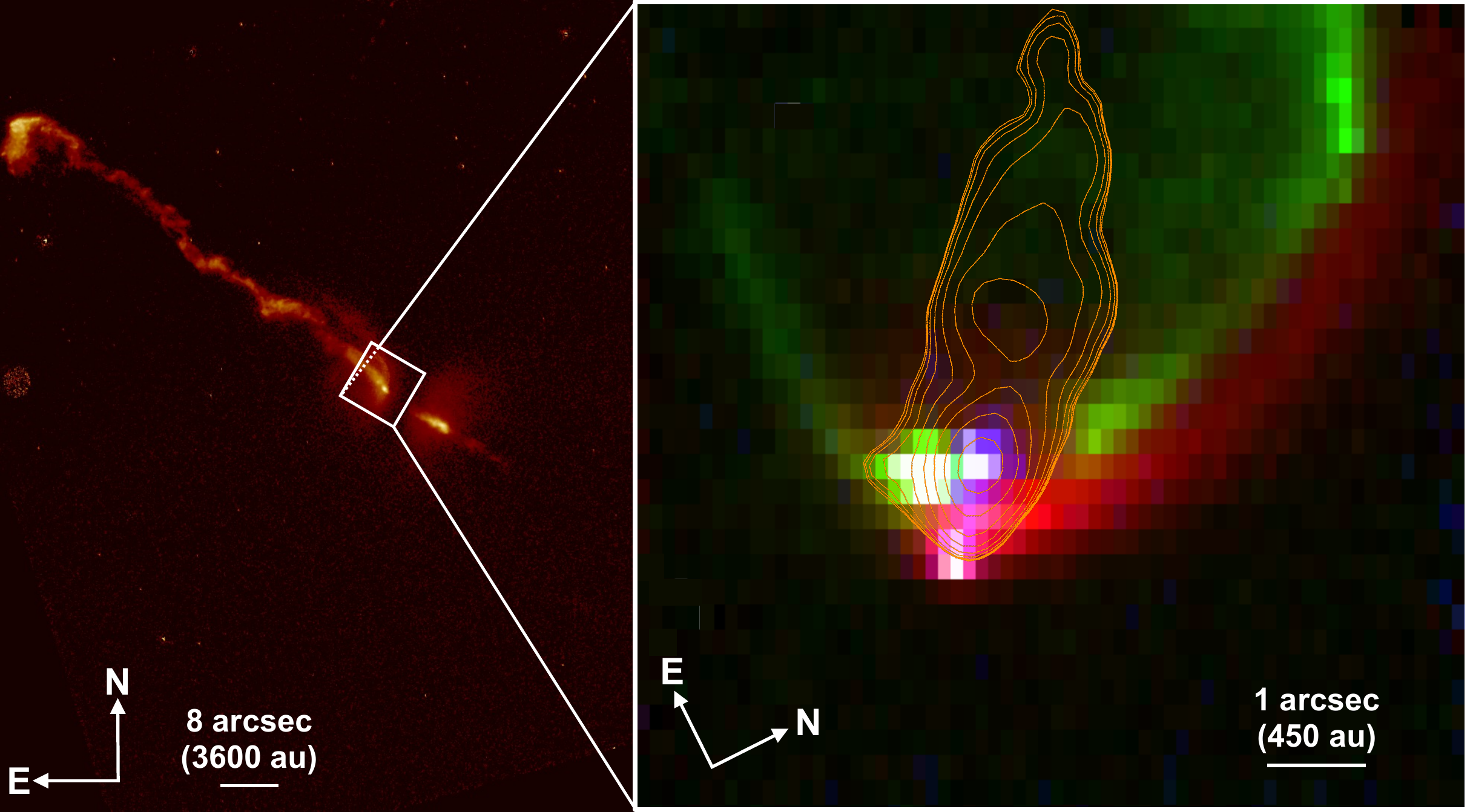}
   \caption{SINFONI three-colour image displaying the outflow cavity and jet compared to HST observations of the jet in [\ion{Fe}{ii}] $\lambda$1.64 $\mu$m emission. Background: Continuum-subtracted [\ion{Fe}{ii}] $\lambda$1.64 $\mu$m emission image of HH 46/47. The observation was obtained using HST WFC3 with the narrowband filter F164N (Program ID: 15178, PI: B. Nisini; \citealt{Erkal2021}). Inset: SINFONI three-colour image displaying the outflow cavity and jet. The red emission traces the integrated continuum emission or scattered light from the outflow driving source. The green emission traces the continuum-subtracted integrated molecular hydrogen (H$_2$) emission displaying the V-shaped outflow cavity. The shocked cavity walls lie within the cavity structure seen in scattered light in continuum emission. There is a bright outflow knot at the base of the left cavity wall. Blue emission traces [\ion{Fe}{ii}] emission, displaying collimated atomic jet emission. The jet culminates in a bright knot or bow shock approximately 1\farcs0 (450 au) from the source position. This knot does not spatially coincide with the outflow knot seen in H$_2$ emission. The inner part of the HST [\ion{Fe}{ii}] $\lambda$1.64 $\mu$m is rotated clockwise by 60\textdegree\ and overplotted in orange contours.}
              \label{3colourimage}%
\end{figure*}

\begin{figure*}
   \centering
   \includegraphics[width=16.1cm, trim= 0cm 0cm 0cm 0cm, clip=true]{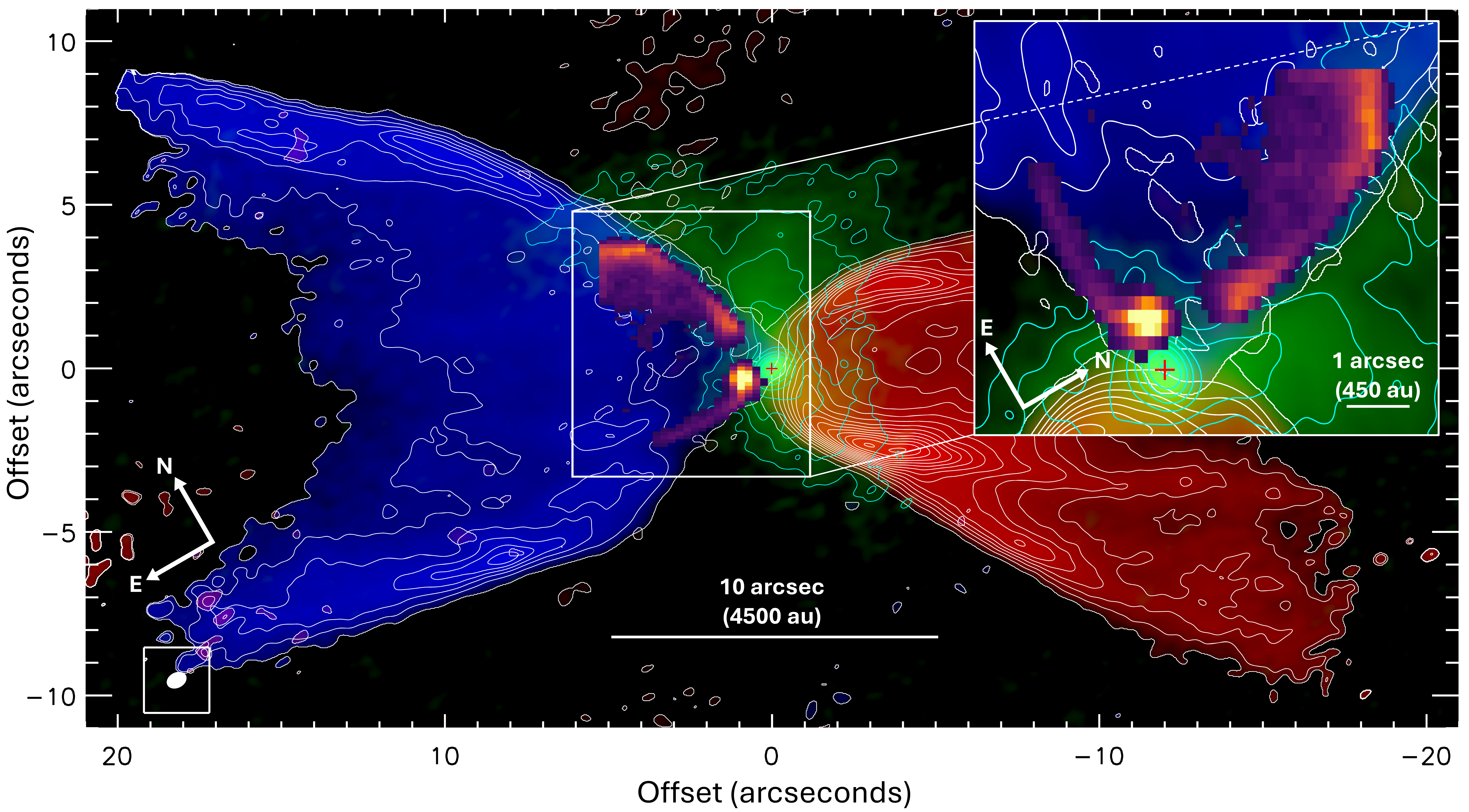}
   \caption{SINFONI H$_2$ emission overplotted on the blueshifted molecular outflow traced by $^{12}$CO (2-1) as observed by ALMA \citep{Zhang2019}. The blue area traces integrated CO emission from $-$35 km~s$^{-1}$ to $-$10 km~s$^{-1}$. The green area highlights the 1.3 mm dust continuum. The red area traces integrated CO emission from +10 km~s$^{-1}$ to +50 km~s$^{-1}$. The shocked cavity walls traced by H$_2$ lie within the cavity structure seen in the CO emission. The red plus indicates the continuum peak position in the SINFONI data. An inset displays a zoomed image of the H$_2$ emission overplotted on the blueshifted CO outflow to better illustrate that the H$_2$ emission is nested within the CO outflow. This inset has been rotated clockwise by 90\textdegree\ to match the convention of figures presented in this paper, where the outflow axis is aligned with the vertical axis. Figure adapted from \cite{Zhang2019}.}
              \label{COlayering}%
\end{figure*}

The cavity seen in continuum emission surrounds the cavity traced by H$_2$ emission, which in turn encompasses the collimated jet revealing a layered asymmetric outflow structure. To investigate this further we compared our SINFONI observations with HST and ALMA observations. Figure \ref{3colourimage} displays the continuum subtracted [\ion{Fe}{ii}] $\lambda$1.64 $\mu$m emission image of HH 46/47, obtained using HST WFC3 with the narrowband filter, F164N (Program ID: 15178, PI: B. Nisini; \citealt{Erkal2021}). The blueshifted jet extends northeasterly and the redshifted jet, southwesterly. The inset shows a three-colour image made from the SINFONI observation of the outflow cavity and jet. Integrated continuum, or scattered light from the outflow driving source, is traced here in red. The continuum and scattered light subtracted, integrated H$_2$ emission is traced here in green. The SINFONI [\ion{Fe}{ii}] continuum subtracted emission, tracing the collimated atomic jet is displayed here in blue. Finally, the inner part of the HST [\ion{Fe}{ii}] $\lambda$1.64 $\mu$m is overplotted in orange contours. For the inset, the HST contours have been rotated to match the 60\textdegree\ rotation that our SINFONI observations underwent. The HST image was aligned with the SINFONI image by coinciding the continuum peak positions, retrieved using 2D Gaussian fitting, located at the apex of the cavity in each image. Figure \ref{COlayering} displays an adapted figure from \cite{Zhang2019} of the blueshifted and redshifted molecular outflows from the HH 46/47 system in $^{12}$CO (2-1) as observed with ALMA. Here, blue traces integrated CO emission from $-$35 km~s$^{-1}$ to $-$10 km~s$^{-1}$, green traces the 1.3 mm dust continuum emission and, red traces integrated CO emission from +10 km~s$^{-1}$ to +50 km~s$^{-1}$. The blueshifted cavity seen in H$_2$ emission is overplotted. The H$_2$ cavity has been rotated anticlockwise by 90\textdegree\ with respect to the SINFONI results presented earlier so that the outflow cavity axis points along the horizontal axis. The continuum peak position in the SINFONI image, marked with a red plus symbol, was aligned with the centre of the 1.3 mm dust continuum in the ALMA image. An inset displays a zoomed image of the H$_2$ emission overplotted on the blueshifted CO outflow to better illustrate that the H$_2$ emission is nested within the CO outflow. This inset has been rotated clockwise by 90\textdegree\ to match the convention of figures presented in this paper, where the outflow axis is aligned with the vertical axis. 

The comparison between SINFONI and HST shows that both sets of observations are consistent. The bending towards the northern cavity wall does not just occur for the $\sim$ 675~au long SINFONI jet but continues out to $>$ 1000~au in the HST jet. Also, the bright knot seen in [\ion{Fe}{ii}] $\lambda$2.047 $\mu$m emission in the SINFONI dataset coincides with a knot traced by the [\ion{Fe}{ii}] $\lambda$1.64 $\mu$m emission from the HST dataset. This is expected as the SINFONI observations took place in November 2018 and HST observations took place in March 2019, just four months later. As the jet knot radial velocity traced by [\ion{Fe}{ii}] $\lambda$2.047 $\mu$m emission is $\sim$ $-$200 km~s$^{-1}$, the jet velocity in the plane of the sky is $\sim$ $-$265 km~s$^{-1}$, or 56 au~yr$^{-1}$, using a jet inclination value of 37\textdegree\ with respect to the plane of the sky from \cite{Hartigan2005}. We therefore expect the jet knot to have moved by just 14 au between observations. The pixel scale of the SINFONI observations is $\sim$ 56 au, meaning the expected offset would not be measurable. The [\ion{Fe}{ii}] $\lambda$1.64 $\mu$m emission displayed in orange contours appears to possess an additional knot coinciding with the H$_2$ knot \textit{A1} seen here in green, suggesting that the knot may also be traced in [\ion{Fe}{ii}] emission. However, due the low S/N, its absence in the SINFONI line to continuum map, and the observation that this knot is not traced by any of the other [\ion{Fe}{ii}] emission lines, we instead attribute this emission to residual unsubtracted continuum emission, scattered light or real emission not detected in these SINFONI observations due to poorer angular resolution. 

The layered structure one sees clearly in Figs. \ref{3colourimage} and \ref{COlayering} is a radial stratification, not only morphological in nature but also in temperature, between hot gas (T$_{H_2}$ $\sim$ $10^3$ K) and the bulk of the surrounding colder material (T$_{CO}$ $\sim$ $10^1$ K). Not only does one see that the H$_2$ emission lies within the continuum emission, but one also sees that the H$_2$ outflow cavity is nested within the blueshifted $^{12}$CO (2-1) cavity. The H$_2$ outflow cavity existing within the CO molecular outflow has been observed in HH 46/47 in \cite{Noriega-Crespo2004}, with different emission lines than those presented here, and more recently in \cite{Nisini2024}, showing that the mid-IR H$_2$ emission is nested within the CO cavity in both the blueshifted and redshifted outflow lobes. At $\sim$ 4\arcsec\ along the outflow axis, the northern H$_2$ cavity wall begins to curl inwards towards the jet axis. This curling of the H$_2$ cavity is also seen in \cite{Nisini2024}, which they interpret as material entrained by the expanding primary jet. Similarly, the eastern cavity wall lies within the CO molecular outflow and appears to possess a narrower opening angle. However, we note that CO opening angle close to the base is narrower than the CO cone farther away from the source, indicating that there is some sudden increase of the CO opening angle, especially evident in the east side at some distance. The inner, narrower CO cone has a morphology more consistent with the H$_2$ emission morphology. This comparison of the different emission tracers reveals a stratification between continuum emission, CO emission, H$_2$ emission, and the jet implying that outflowing material from the star-disc system is spatially layered as radial distance from the jet axis increases. The highest velocity, highest temperature, and most strongly collimated emission is located at the centre of the outflow axis and surrounding it is the slower moving, cooler, wide-angled molecular component of the outflow. This onion-like layered flow structure with stratification of emission in velocity, temperature, and chemistry has been observed in some cases \citep{Delabrosse2024}, and is predicted by MHD disc wind models \citep{Ferreira1997, Panoglou2012, Bai2016, Zhu&Stone2018, Wang2019}. In these models, material is accelerated along flow streamlines, poloidal magnetic field lines anchored in the rotating disc. Hot, high-velocity emission travel along streamlines originating in the very inner disc region, whilst cooler, lower velocity emission travel along streamlines originating further out in the disc. This stratified structure can be identified observationally when tracing the disc wind directly or tracing the winds interaction with surrounding ambient material. In this case, the CO is thought to be entrained material, tracing material located outside the cavity wall, whilst the H$_2$ emission is likely tracing the interaction region that put the entrained material into motion. The origin of such an interaction can be explained by an MHD disc wind, but can also be produced by the jet alone. Consequently, we cannot conclusively attribute the onion-like stratified structure exclusively to an MHD disc wind.

%The background in figure \ref{3colourimage} }

%The lower velocity molecular outflow traced by H$_2$ is less strongly collimated and surrounds the high-velocity jet is a cone-like 3-dimensional morphology. The even lower velocity molecular outflow traced by CO is even less strongly collimated and appears to surround the H$_2$ outflow cavity. 

The H$_2$ cavity is likely an interaction region between some wide-angled component of the outflow, be it an uncollimated or poorly collimated wind, or an expanding shell of material produced by a wide-angled wind, and the surrounding ambient material, or by the stacking of successive jet bowshocks from a collimated, variable inner jet propagating into a density-stratified surrounding medium \citep{Rabenanahary2022}. It is possible that we are seeing nested cavities traced in H$_2$ similar to those seen in cooler CO emission caused by multiple expanding shells. These expanding shells of material have been observed in the blueshifted and redshifted outflows of HH 46/47, traced by $^{12}$CO emission \citep{Zhang2019}. These shells, which are highly coherent in space and velocity, have an parabolic morphology. The origin of these outflowing shells can be explained by entrainment of ambient material by an episodic wide-angle wind \citep{Li&Shu1996, Lee2000} , which produces radially expanding parabolic shells, or by successive jet bowshocks. The general morphology of the H$_2$ cavity that lies within the cavity traced by continuum emission could be produced by expanding shells of material, interior to the shells identified in CO by \cite{Zhang2019}. Furthermore, the emission peaks labelled \textit{A1} and \textit{A2} could potentially be tracing shocked interaction regions between a more evolved shell, traced by the outer cavity walls, and a smaller scale, asymmetric H$_2$ shell nested within, produced by a subsequent outburst of the wide-angled wind or a successive jet bowshock. The arc-shaped morphological component labelled \textit{A3} could also be associated with a cavity or shocked region created by an expanding shell. In terms of the shock conditions in the H$_2$ cavity one can look at the spectra presented in Figs. \ref{Stellarspec} \& \ref{HH4647spectra}. When one compares the intensities of H$_2$ 1-0 emission lines, those originating from the $v$ = 1-0 vibrational transition such as the H$_2$ 1-0 S(1) line at 2.12 $\mu$m, are relatively much brighter than the H$_2$ 2-1 lines, those originating from the $v$ = 2-1 vibrational transition such as the H$_2$ 2-1 S(1) line at 2.25 $\mu$m. This indicates that the shocks in the H$_2$ cavity are low velocity and non-dissociative. The shock velocities are not high enough to dissociate the molecular hydrogen, or heat the gas to a sufficient temperature to significantly populate higher vibrational levels such as $v$ = 2 or higher.

It is clear that both the outflow cavity traced by H$_2$ emission and the atomic jet traced by [\ion{Fe}{ii}] and Br-$\gamma$ emission are asymmetric about the CO outflow direction. The emission region of the left side of the cavity, that is, the eastern cavity wall, traced by H$_2$ is spatially more compact when compared to the right side, the northern cavity wall. The right side of the cavity also extends further from the driving source, both vertically and laterally. If the surrounding medium on the right side of the cavity side is denser than that of the left side, then this would naturally lead to brighter shocked wall regions on the right side of the cavity. The suggestion that the surrounding ambient material is denser on the right side of the cavity is further supported by the morphology of the scattered light emission seen in Fig. \ref{Morphologyfig}. Here, the cavity as seen in scattered light is highly asymmetric about the CO outflow axis with the northern arm extending much further from the driving source than the eastern arm. This further suggests a dissymmetry in the distribution of the surrounding ambient material. This inhomogeneous ambient material is further testified by the general morphology of the reflection nebula seen at optical wavelengths that extends mostly in the northern direction \citep{Heathcote1996}. Furthermore, in Fig. \ref{COlayering}, the 1.3 mm dust continuum is clearly brighter on the northern side, supporting this argument. As mentioned before, there is some sudden increase of the CO opening angle, especially evident in the east side at between 2\arcsec\ and 3\arcsec\, additionally reinforcing this argument. The northern side of the outflow is facing the cloud edge that the blueshifted outflow is emanating from, which may be shaped or compressed by radiation from outside the cloud. This may provide a rationale for why the ambient material on the northern side appears denser. There also exists and offset between the average PA of the atomic jet, 52.3\textdegree\ from \cite{Erkal2021}, and the central axis of the wide-angled CO outflow, $\sim$ 60\textdegree\ from \cite{Zhang2019}. This difference between the jet PA and molecular outflow PA can be interpreted as being due to an inhomogeneous surrounding medium, with the eastern cavity wall expanding quicker than the northern cavity wall and therefore shifting the central axis of the CO outflow to a larger PA than the jet PA. 

We see that the atomic jet is deflected away from the left side of the cavity, tending towards the northern wall. Also, the jet knot traced by [\ion{Fe}{ii}] and Br-$\gamma$ emission does not coincide with either peak traced at the bases of the cavity walls seen in H$_2$ labelled \textit{A1} and \textit{A2}. The bending of the atomic jet towards the northern cavity wall is consistent with the [\ion{Fe}{ii}] $\lambda$1.64 $\mu$m emission from the HST observations. This bending of the atomic jet is indicative of jet wiggling, which is clearly seen in large scale observations of the jet \citep{Reipurth2000, Erkal2021}. Jet wiggling can be induced by an orbiting companion to the outflow driving source \citep{Murphy2021}. These axial asymmetries seen in the blueshifted outflow, both in the atomic jet and the wide-angled outflow and cavity may also arise due to the influence of two outflows being driven by each of the components of the binary system \citep{Nisini2024}. The difference between the jet PA and molecular outflow PA can also be interpreted as being due to the secondary jet pushing on the eastern cavity wall, again shifting the central axis of the CO outflow to a different PA when compared to the jet PA.

\subsection{Origin of the H$_2$ transverse velocity gradient} \label{sec: Origin_velgrad}

The V-shaped, parabolic cavity observed in the integrated H$_2$ emission in Fig. \ref{velgrad} displays a clear transverse velocity gradient of approximately $\sim$ 10 km~s$^{-1}$ between the eastern and northern pointing cavity walls (see Table \ref{Table:Velocity difference}). A transverse velocity difference, lower in magnitude, is also present in $^{12}$CO emission in the same sense as the velocity difference presented here (2024, priv. comm Y. Zhang). Whilst the spectral resolution of the SINFONI dataset is R = 4000 or 2.45 Å, which translates to a spectral sampling of 38 km~s$^{-1}$ at 1.95 $\mu$m and 30 km~s$^{-1}$ at 2.45 $\mu$m, we can be confident in a velocity difference of $\sim$ 10 km~s$^{-1}$ between either side of the cavity due to our wavelength recalibration performed using the OH lines from the upper atmosphere and discussed in Sect. \ref{WavelengthRecal}. We estimate a 1$\sigma$ uncertainty of $\pm$ 1.0 km~s$^{-1}$ to $\pm$ 3.0 km~s$^{-1}$ associated with this velocity centroid map, which encompasses absolute and relative errors associated with Gaussian centroid accuracy and the OH wavelength calibration.

Initially one may interpret that this velocity gradient one sees is due to the anticlockwise rotation of the H$_2$ cavity, with the eastern cavity wall approaching at a higher velocity with respect to the northern cavity wall. This apparent rotation may be tracing a rotating disc wind directly \citep{Zhang2016}, or tracing the rotation of entrained material, where the angular momentum present in the envelope is provided for by the jet \citep{Gaudel2020}. This rotation would support the MHD driven wind model, as one expects a rotation of the outflow if the wind is removing angular momentum from the star-disc system. Rotating cavities have been observed in early to late stages of low mass star formation \citep{Bjerkeli2016, Tabone2017, Zhang2018, Louvet2018, deValon2022, Launhardt2023}. In order to investigate the plausibility of this velocity gradient being caused by a rotation of the cavity, we assumed that the velocity gradient was purely due to rotational motion and calculated the specific angular momentum of this apparent rotation at different heights along the outflow. We derive the velocity differentials from the H$_2$ velocity centroid map provided in Fig. \ref{velgrad}, by extracting a transverse PV cut at a specific height along the outflow. Gaussian profiles are then fit along the velocity axis to both of the emission peaks corresponding to each side of the H$_2$ cavity, retrieving the centroid velocity of each side of the H$_2$ cavity, as described in Sect. \ref{Sec: H2 centroid map}. The rotational radius \textit{r} is estimated, assuming the rotation is axisymmetric, as half of the distance between the two emission peaks corresponding to each side of the H$_2$ cavity. The uncertainties associated with the centroid velocities are quadrature summations of the Gaussian centroid error and the uncertainty obtained with wavelength recalibration, as discussed in Sect. \ref{WavelengthRecal}. The specific angular momentum ($J = r \times v_{\mathrm{rot}}$) of a rotation or orbit is defined as the angular momentum associated with the rotation divided by the systems mass, in other words, the angular momentum per unit mass. Assuming that the flow is axisymmetric we can calculate the specific angular momentum as follows: 

% \begin{equation} \notag
%     j = \frac{J}{M} = \frac{I\Omega}{M} = \frac{Mr^2\Omega}{M} = r^2\Omega
% \end{equation}
\begin{equation} \label{eq: momentum}
    J(r)\ = \ r \times v_{\mathrm{rot}}(r) \ = \ r \times \frac{v_{\mathrm{left}}\: - \: v_{\mathrm{right}}}{2 \, \mathrm{sin}(i)}
,\end{equation}
where $v_{\mathrm{rot}}(r)$ is the rotational velocity and $i$ is the angle of inclination of the flow with respect to the line of sight, $i$ = 53\textdegree. Using Eq. \ref{eq: momentum}, we calculated the specific angular momentum of the apparent cavity rotation at heights along the outflow in the plane of the sky, of $z$ = 450 au, 675 au, and 900 au ($z$ = 1\farcs0, 1\farcs5, 2\farcs0). The specific angular momentum results are presented in Table \ref{Table:SpecificAngularMomentum}. We calculate specific angular momenta values of $J(r)$ = 3217 $\pm$ 658 km~s$^{-1}$au, at 1\arcsec\ offset from the central source, to $J(r)$ = 3867 $\pm$ 1125 km~s$^{-1}$au, at 2\arcsec\ offset.

\begin{table*}
\caption{Specific angular momenta calculations of H$_2$ cavity rotation at various heights along the outflow.} 
\label{Table:SpecificAngularMomentum}
    \centering
    \begin{tabular}{cccccccc}
    \hline\hline
        $z$ (au) & $v_{\mathrm{left}}$ (km~s$^{-1}$) & $v_{\mathrm{right}}$ (km~s$^{-1}$) &$v_{\mathrm{rot}}$ (km~s$^{-1}$) & $r$ (au) & $r$ (10$^{-3}$ pc) & $J(r)$ (km~s$^{-1}$au) & $J(r)$ (10$^{-2}$ km~s$^{-1}$pc) \\ \hline
        450 & 31.3 $\pm$ 1.5 & 17.3 $\pm$ 2.5 & 8.8 $\pm$ 1.8 & 365.6 & 1.8 & 3217 $\pm$ 658 & 1.6 $\pm$ 0.3 \\ %\cmidrule(l r){1-8}
        675 & 25.1 $\pm$ 2.0 & 13.4 $\pm$ 1.4 & 7.3 $\pm$ 1.5 & 506.3 & 2.5 & 3696 $\pm$ 759 & 1.8 $\pm$ 0.4 \\  %\cmidrule(l r){1-8}
        900 & 24.0 $\pm$ 2.1 & 15.2 $\pm$ 1.4 & 5.5 $\pm$ 1.6 & 703.1 & 3.4 & 3867 $\pm$ 1125 & 1.9 $\pm$ 0.6 \\  %\cmidrule(l r){1-8}
        \hline
    \end{tabular}
    %\captionsetup{justification=centering}
    %\captionsetup{tableposition=top}
\end{table*}

In the case that the H$_2$ velocity gradient is directly tracing an outflow rotation we can compare to previous outflow rotation observations. \cite{Bjerkeli2016} measure the specific angular momentum, in $^{12}$CO emission, of the rotating outflow associated with the TMC1A protostellar system to be less than 200 km~s$^{-1}$au, using \textit{r} values an order of magnitude lower than those used here. \cite{Zhang2018} measure the mean specific angular momentum, in C$^{18}$O emission, in the rotating NGC 1333 IRAS 4C outflow to be 100 km~s$^{-1}$au. \cite{Louvet2018} measure the specific angular momentum, in $^{12}$CO emission, of the rotating HH 30 outflow to be $\sim$ 40 km~s$^{-1}$au. In \cite{deValon2022}, they measure the specific angular momentum, in $^{12}$CO emission, of the rotating outflow from DG Tauri B to be $\sim$ 65 km~s$^{-1}$au. \cite{Launhardt2023} measure the specific angular momentum, in $^{12}$CO emission, of the rotating outflow from a young T Tauri star in the CB 26 Bok globule to be less than 200 km~s$^{-1}$au. \cite{Zhang2016} estimated the specific angular momentum in the equatorial plane of the envelope surrounding HH 46/47 in $^{13}$CO and C$^{18}$O emission. They found specific angular momentum values for the rotation of $J$ = 450 - 540 km~s$^{-1}$au ($J$ =  2.2 $\times$ 10$^{-3}$ to 2.6 $\times$ 10$^{-3}$ km~s$^{-1}$pc), an order of magnitude lower than the specific angular momenta estimated here. As this rotation is measured in the equatorial plane, it is indicative of an upper limit for the specific angular momentum in this system. In the case that the H$_2$ velocity gradient is instead tracing the rotation of entrained material, where the angular momentum present in the envelope is provided for by the jet we can compare with \cite{Gaudel2020}. They measure the specific angular momentum of the midplanes of circumstellar environments of a sample of class 0/1 protostellar envelopes. In their sample, the maximum specific angular momenta values, at similar \textit{r} values presented here, are 2 $\times$ 10$^{-3}$ km~s$^{-1}$pc ($\sim$ 400 km~s$^{-1}$au), an order of magnitude below what we measure here. 

The specific angular momenta calculated here at various heights along the outflow cavity are significantly larger than what has been seen before for rotating outflow cavities based on either rotational motion traced directly by the outflow or by rotating entrained material. Furthermore, the apparent anticlockwise rotation we see here is opposite to the clockwise rotation of the flattened structure around the central source perpendicular to the outflow axis traced by $^{13}$CO and C$^{18}$O from \cite{Zhang2016}. Whilst this flattened structure has a size of approximately 10\arcsec\ (4500 au) across, which is much larger than what is expected for a rotationally supported Keplerian disc, it is indicative of the disc rotation direction as the sense of rotation remains consistent as the beam size, in $^{13}$CO and C$^{18}$O, is scaled down to its smallest size, 1\farcs5 (675 au). They conclude that the structure is a rotating envelope that feeds the accretion disc. Although some jets present transverse velocity gradients in the opposite sense to the disc rotation \citep{Cabrit2006, Louvet2016}, it is unclear if these gradients (of a few kilometres per second and variable in time) truly trace counter-rotation of the jet, or if they are contaminated by other effects such as jet precession or shock asymmetries. The large specific angular momenta calculated here and the fact that the apparent rotation is opposite to the sense of the disc rotation lead us to rule out the interpretation that the velocity gradient observed in the H$_2$ cavity is entirely tracing a rotation signature and other effects must be considered. 

Another possible explanation for the velocity gradient could be ascertained from the observation that the cavity seen in H$_2$ is clearly asymmetric about the outflow direction. If the surrounding medium in the region of the northern cavity wall is denser than that in the region around the eastern cavity wall, as discussed in Sect. \ref{sec: layeringOfEmission}, then the outflowing material, or expanding shells of material rather than a wide-angled wind, would encounter and impact denser quiescent material in the region of the northern cavity wall causing the outflow or shells to decelerate more quickly. This could explain why the northern wall of the cavity displays lower radial velocities when compared to that seen in the eastern wall. The denser quiescent material at the right side of the cavity could naturally explain the velocity gradient one sees in the H$_2$ emission. The production of transverse velocity gradients due to dissymmetry in density of surrounding ambient material has been discussed by \cite{DeColle2016}.

Lastly, the velocity gradient seen in the cavity traced by H$_2$ emission could be that the dominating outflow driver for the northern and eastern cavity walls originate from different sources. The driving source of the HH 46/47 outflow system is known to be a binary source \citep{Reipurth2000}. \cite{Nisini2024} suggest that the bright H$_2$ knot seen at the base of the eastern cavity wall, \textit{A1}, is a expanding bubble ejected by the secondary component of the binary system. This could imply that the emission one sees in the eastern wall of the H$_2$ cavity could have been strongly influenced by this secondary outflow driver. The outflow from this secondary source in the binary system could be strongly contributing to the shocked H$_2$ emission in the eastern wall of the cavity, potentially increasing it's radial velocity in comparison to the northern wall. This is further testified by the H$_2$ knot \textit{A1} possessing the highest blueshifted radial velocity in the velocity centroid maps provided in Fig. \ref{velgrad}.

We would like to emphasise that velocity gradient signatures observed transverse to the outflow direction can easily be misinterpreted as being due to rotational motion, and this interpretation is further motivated by current MHD disc wind models. However, other effects, such as ambient material density variations and the presence of multiple outflows, as discussed above can produce transverse velocity gradient signals. In addition, further effects such as kink instabilities \citep{Staff2014} and twin-jet structures \citep{Soker2022}, to name a few, can reproduce transverse velocity gradients. We investigated other interpretations of the observed velocity gradient due to our specific angular momentum values being significantly larger than that seen before for rotating outflow cavities, and due to the apparent counter-rotation of the outflow with respect to the conjectured disc rotation direction. However, other effects that can produce similar velocity gradient signatures must be considered in cases where specific angular momentum estimates are as expected and apparent rotation direction is in the same sense as the disc.  

\section{Summary and conclusions} \label{sec: Conclusions}
We have investigated the morphology and kinematics of the launching region of the blueshifted HH 46/47 outflow emanating from the Class I binary system HH 46 IRS with SINFONI K-band observations. We achieved an angular resolution of 0\farcs81 and a precision in velocity centroids, in H$_2$ 1-0 S(1) emission, of $\pm$ 1 to 3 km~s$^{-1}$. These observations allowed us to study both the base of the atomic jet traced by [\ion{Fe}{ii}] and Br-$\gamma$ emission and the wide-angled molecular hydrogen outflow. We summarise our main results as follows: 

\begin{itemize}

    \item[--] We have presented the morphology of the NIR continuum, H$_2$ 1-0 S(1), [\ion{Fe}{ii}] $\lambda$2.047 $\mu$m and Br-$\gamma$ emission. The continuum emission displays an approximately parabolic and asymmetric reflection nebula that we interpret as a boundary region between the outflow and the ambient material. Where the outflow has carved a wide-angled cavity in the quiescent material. We recovered a point source at the apex of this cavity that we establish as the source position. The H$_2$ emission displays a striking V-shaped morphology tracing the limb-brightened cavity walls of the outflow. We infer that the cavity structure we observe has a 3D inverted-conical shape surrounding the atomic jet. The forging of this outflow cavity, which we determine to be an interaction region between the outflow and the surrounding quiescent material, may be due to some wide-angled wind or expanding shells of material produced by an episodic wide-angled wind or by the stacking of successive jet bowshocks from a collimated variable inner jet propagating into a density-stratified surrounding medium. The morphological components labelled \textit{A1}, \textit{A2}, and \textit{A3} may arise due to multiple expanding shells, with a smaller scale shell associated with the shocked emission peaks \textit{A1} and \textit{A2} expanding within a slightly larger scale shell associated with the outer cavity walls produced by an earlier outburst of the wide-angled wind or by a successive jet bowshock. The high-velocity [\ion{Fe}{ii}] and Br-$\gamma$ emission traces the base of the atomic jet that culminates in a jet knot at approximately 1\farcs0 or 450 au from the source position. These conclusions are similar to those of \cite{Nisini2024}.
\\
    \item[--] We find radial stratification of the outflow in different emission tracers, which produces an onion-like flow structure, as discussed in Sect. \ref{sec: layeringOfEmission}. The slow-moving wide-angled molecular outflow cavity traced in H$_2$ emission surrounds the base of the hotter high-velocity atomic jet traced in [\ion{Fe}{ii}] and Br-$\gamma$ emission. The H$_2$ outflow cavity lies within the reflection nebula seen in continuum emission, and it extends less laterally and has a narrower opening angle. The H$_2$ outflow cavity is also nested within the molecular outflow traced in $^{12}$CO (2-1), and it appears to curl inward towards the outflow axis at a distance along the outflow of approximately 4\arcsec.
\\   
    \item[--] The reflection nebula seen in continuum emission, the H$_2$ outflow cavity, and the atomic jet traced in [\ion{Fe}{ii}] and Br-$\gamma$ are all asymmetric about the CO outflow direction. This is discussed in Sect. \ref{sec: layeringOfEmission}. We provide explanations for the origin of the H$_2$ outflow cavity asymmetries, including density variations in the surrounding ambient material and the influence of multiple outflows. It is plausible that both components of the binary source drive the outflows. We observed bending of the atomic jet, which emanates from the source position and tends towards the northern cavity wall, and it is indicative of the jet wiggling that we have observed in this jet at larger scales. 
\\    
    \item[--] We used PV diagrams, as seen in Sect. \ref{kinematics}, to investigate the kinematics of the atomic jet and the H$_2$ outflow cavity. The radial velocity of the atomic jet emission traced by [\ion{Fe}{ii}] and Br-$\gamma$ peaks at $\sim$ $-$210 km~s$^{-1}$, which is consistent with previous measurements. The Gaussian centroid velocity of the atomic jet knot at an offset of 1\arcsec\ from the source position is $v_{\mathrm{LSR}}$ = $-$208.7 $\pm$ 3.6 km~s$^{-1}$ in [\ion{Fe}{ii}] emission and $v_{\mathrm{LSR}}$ = $-$211.9 $\pm$ 3.5 km~s$^{-1}$ in Br-$\gamma$ emission. A PV diagram taken transverse to the outflow axis was produced to investigate the kinematics of the H$_2$ emission. The radial velocity of the H$_2$ emission is significantly lower than that of the atomic jet emission, peaking around $\sim$ $-$15 km~s$^{-1}$ to $-$30 km~s$^{-1}$. We observed two emission peaks located at the base of each cavity wall on either side of the source position, labelled \textit{A1} and \textit{A2}. The emission peak from the base of the eastern cavity wall (\textit{A1}) peaks at a slightly higher radial velocity ($-$33.9 $\pm$ 1.4 km~s$^{-1}$) than the emission peak (\textit{A2}) from the base of the northern cavity wall ($-$16.9 $\pm$ 1.5 km~s$^{-1}$).
\\
    \item[--] We confirm the presence of a velocity gradient in the outflow cavity, transverse to the outflow axis, traced in H$_2$. We generated a velocity centroid map to probe this observed velocity gradient. The magnitude of this velocity difference is $\sim$ 10 km~s$^{-1}$ and exists between the northern and eastern pointing cavity walls (see Table \ref{Table:Velocity difference}). We have discussed some possible explanations for the origin of this velocity gradient (Sect. \ref{sec: Origin_velgrad}). We ruled out outflow rotation as the sole origin due to large specific angular momenta values, $J(r)$ = 3217 $\pm$ 658 to 3867 $\pm$ 1125 km~s$^{-1}$au, calculated from 1\arcsec\ to 2\arcsec\ offset from the central source and the fact that the sense of apparent rotation we detected is opposite to that of the CO envelope emission. We instead favour a dissymmetry in the ambient material surrounding the outflow or the presence of multiple outflows. 
    
\end{itemize}

This paper displays the wealth of information that can be extracted from IFU observations of protostellar outflows on intermediate scales in the NIR and highlights the morphological and kinematical complexity that needs to be considered when interpreting various aspects of outflows at these scales. In particular, we want to emphasise that radial velocity gradients observed transverse to the outflow axis can be misinterpreted as rotation signatures when other effects, such as those we discussed, can produce equivalent transverse velocity gradients. 

\begin{acknowledgements}
      This work supported by the John \& Pat Hume Doctoral Scholarship at Maynooth University (MU) and the Travelling Doctoral Studentship from the National University of Ireland (NUI). BN acknowledge support from the Large Grant INAF 2022 “YSOs Outflows, Disks and Accretion: towards a global framework for the evolution of planet forming systems (YODA)” and from PRIN-MUR 2022 20228JPA3A “The path to star and planet formation in the JWST era (PATH)”.
\end{acknowledgements}


%\begin{thebibliography}
\bibliographystyle{aa} 
\bibliography{bibliography}

%   \bibitem[Baker(1966)]{baker} Baker, N. 1966,
%       in Stellar Evolution,
%       ed.\ R. F. Stein,\& A. G. W. Cameron
%       (Plenum, New York) 333

%    \bibitem[Balluch(1988)]{balluch} Balluch, M. 1988,
%       A\&A, 200, 58

%    \bibitem[Cox(1980)]{cox} Cox, J. P. 1980,
%       Theory of Stellar Pulsation
%       (Princeton University Press, Princeton) 165

%    \bibitem[Cox(1969)]{cox69} Cox, A. N.,\& Stewart, J. N. 1969,
%       Academia Nauk, Scientific Information 15, 1

%    \bibitem[Mizuno(1980)]{mizuno} Mizuno H. 1980,
%       Prog. Theor. Phys., 64, 544
   
%    \bibitem[Tscharnuter(1987)]{tscharnuter} Tscharnuter W. M. 1987,
%       A\&A, 188, 55
  
%    \bibitem[Terlevich(1992)]{terlevich} Terlevich, R. 1992, in ASP Conf. Ser. 31, 
%       Relationships between Active Galactic Nuclei and Starburst Galaxies, 
%       ed. A. V. Filippenko, 13

%    \bibitem[Yorke(1980a)]{yorke80a} Yorke, H. W. 1980a,
%       A\&A, 86, 286

%    \bibitem[Zheng(1997)]{zheng} Zheng, W., Davidsen, A. F., Tytler, D. \& Kriss, G. A.
%       1997, preprint
%\end{thebibliography}

\begin{appendix}

\section{OH wavelength calibration map}
\label{AppSect: OH calibration}
\begin{figure}
    \centering
    \includegraphics[width=8.2cm, trim= 0cm 0cm 0cm 0cm, clip=true]{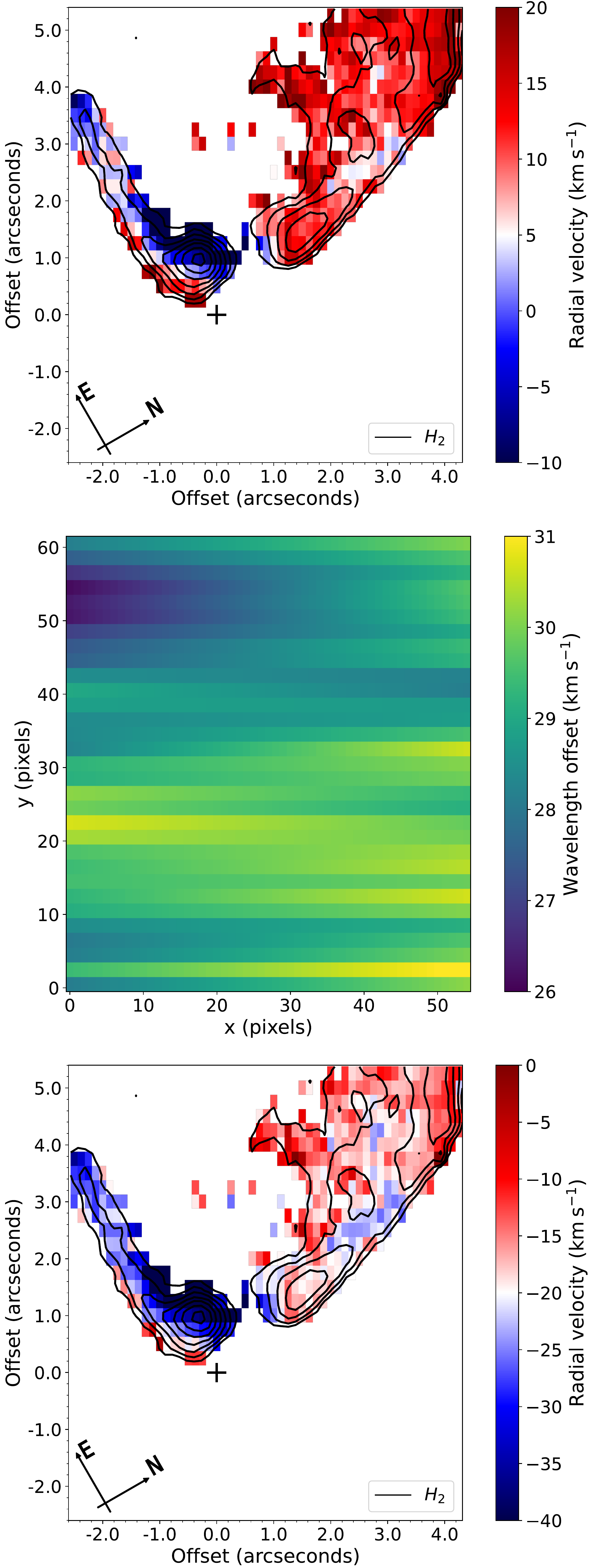}
    \caption{Velocity centroid maps of H$_2$ emission before (top), and after (bottom), the wavelength recalibration map (centre), derived from the OH emission lines, is applied. Notice the change in velocity scales used for the velocity centroid maps. The source position is represented with a black plus symbol in the velocity centroid maps. Contour levels for both begin at 3$\sigma$ of the background emission and increase by factors of 1.5.}

    \label{OHmap}
\end{figure}

Figure \ref{OHmap} illustrates the effect of the OH line wavelength recalibration on the H$_2$ emission velocity centroid maps. The magnitude of the wavelength-velocity correction, shown in the central panel, varies between $\sim$ $-$26 km~s$^{-1}$ and $-$30 km~s$^{-1}$ and varies most significantly in the vertical direction. This wavelength correction map was applied to the SINFONI data cube around the H$_2$ 1–0 S(1) emission line at 2.12 $\mu$m, effectively obtaining a more accurate wavelength calibration than that offered with the standard arc lamp wavelength calibration.

\FloatBarrier

%\clearpage
\section{Uneven slit effect correction}
\label{AppSect: Uneven slit}

To correct for uneven slit illumination, we followed the method outlined in \cite{Agra-Amboage2014, Erkal2021_b}. First, we modelled the distribution of incoming light by spectrally integrating the emission line of interest and the local continuum emission either side of the emission line. We then estimated the brightness centroid and its displacement with respect to the centre of the slitlet at each spaxel in the data cube. The spurious wavelength shift, expressed in terms of a velocity shift, is calculated using the adapted formula, derived by \cite{Marconi2003}: 

\begin{equation} \label{Eq: uneven slit}
    \Delta v(x_0,y_0) = \frac{\delta u}{\delta y} \times \frac{\iint I_{\mathrm{mod}}(u,v) \times (v-y_0)~\mathrm{d}u~\mathrm{d}v}{\iint I_{\mathrm{mod}}(u,v)~\mathrm{d}u~\mathrm{d}v}
\end{equation}

\begin{figure}
    \centering
    \includegraphics[width=8.2cm, trim= 0cm 0cm 0cm 0cm, clip=true]{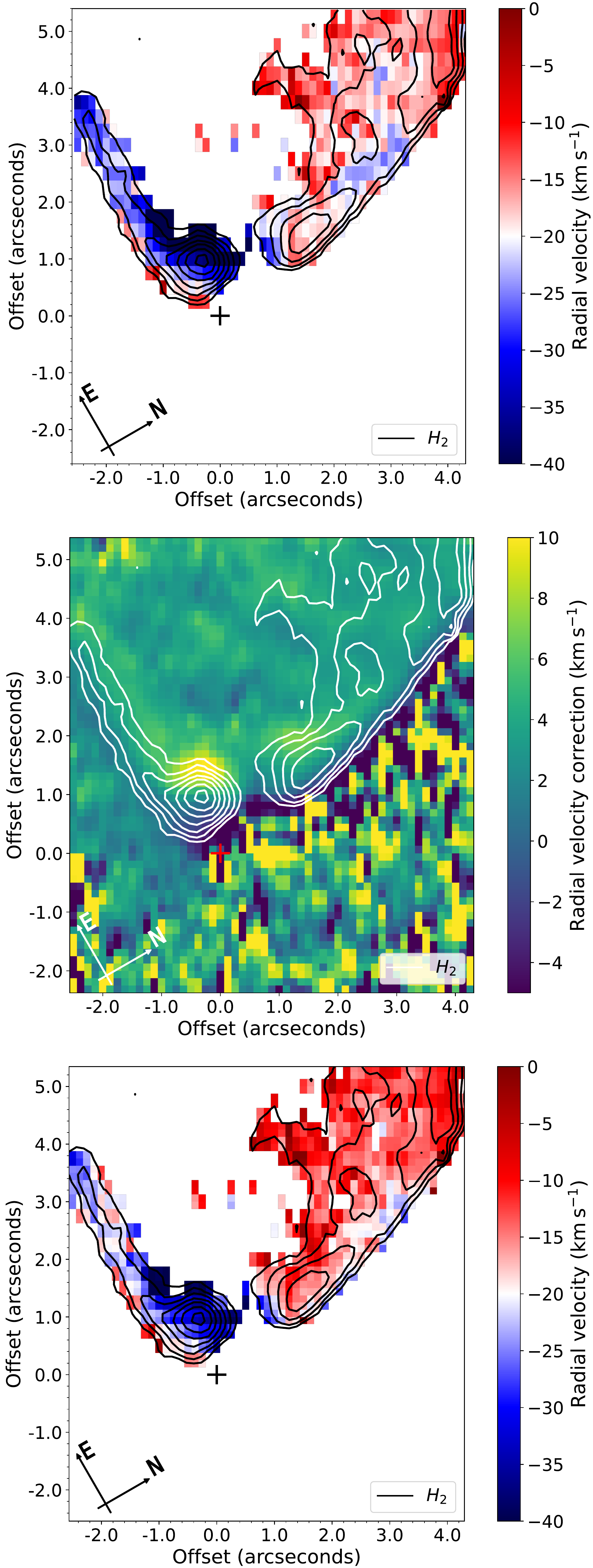}
    \caption{Velocity centroid maps of H$_2$ emission before (top), and after (bottom), the wavelength correction map required to amend illusory wavelength-velocity shifts due to uneven slit illumination (centre), is applied. In the central figure H$_2$ emission is overplotted in white contours. The continuum peak position is marked with a red plus symbol. Contour levels for the 3 panels begin at 3$\sigma$ of the background emission and increase by factors of 1.5.}
    \label{unevenslitmap}
\end{figure}

Here, $\delta u$ is the spectral pixel sampling, $\delta y$ is the width of the slitlet, and $I_{\mathrm{mod}}$ is the modelled light intensity integrated over a single spaxel of detector coordinates $(x_0, y_0)$. Figure \ref{unevenslitmap} displays the 2D wavelength correction map computed around the H$_2$ $\lambda$2.122 $\mu$m emission line. Velocity corrections here vary from $-$5 km~s$^{-1}$ to +10 km~s$^{-1}$. As expected, the velocity corrections have a strong variation in the vertical, or dispersion direction due to the nature of the horizontal slicing mirrors present in the SINFONI instrument. The effect due to uneven slit illumination is more prominent in brighter emission and more compact emission.

\FloatBarrier

\section{Effect of scattered emission removal}
\label{AppSect: Scattered emission removal}
\begin{figure}[h!tbp]
    \centering
    \includegraphics[width=8.2cm, trim= 0cm 0cm 0cm 0cm, clip=true]{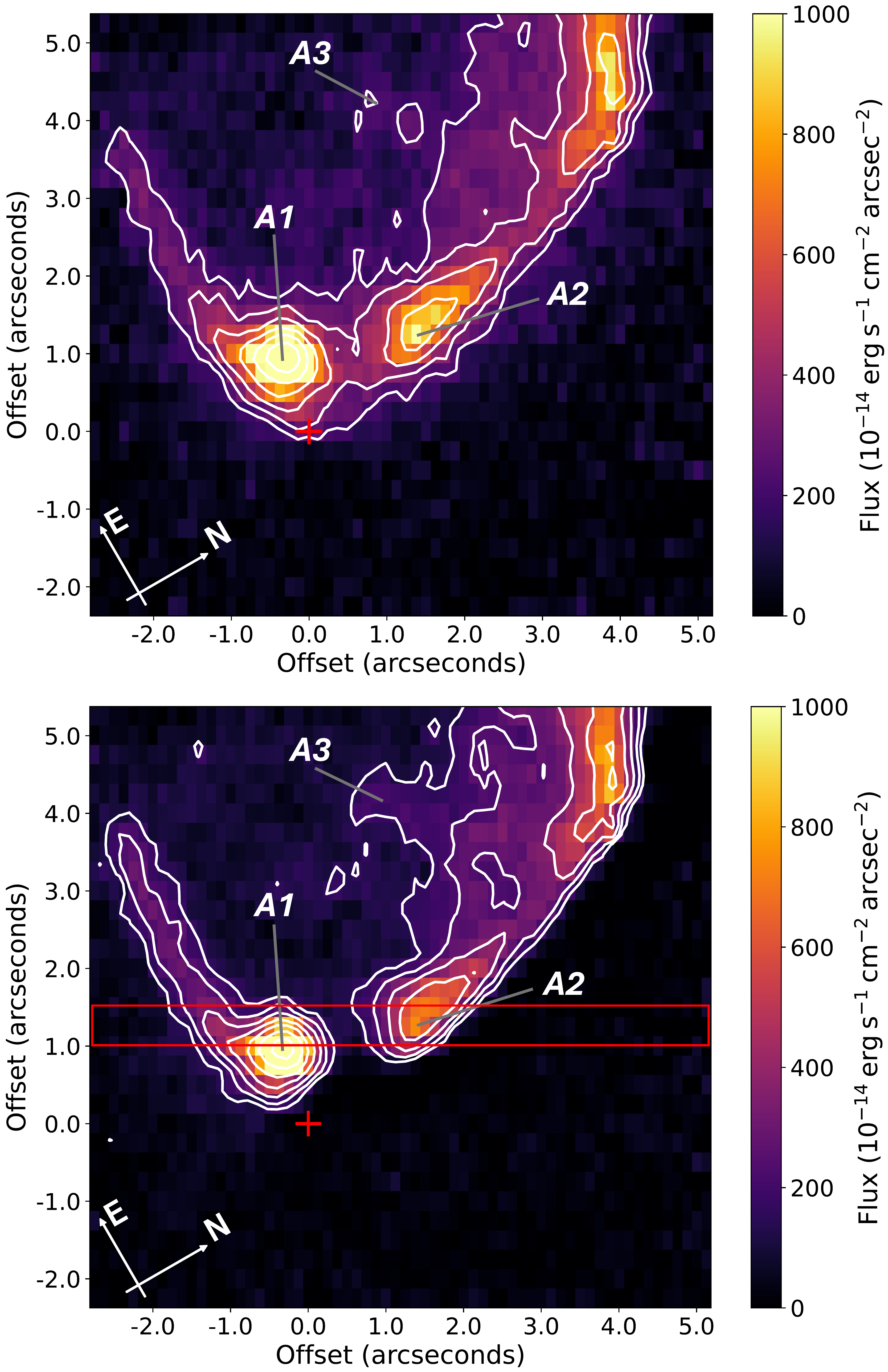}
    \caption{Integrated H$_2$ line emission before (top) and after (bottom) the scattered emission removal routine is preformed. The source position is represented with a red plus symbol. Contour levels for both begin at 3$\sigma$ of the background emission and increase by factors of 1.5. The red rectangle in the bottom panel displays the position and width of the pseudo-slit used to extract the transverse PV diagram seen in Fig. \ref{pvtransverse}.}

    \label{Scattered_emission_comparison}
\end{figure}

Figure \ref{Scattered_emission_comparison} displays the integrated H$_2$ 1-0 S(1) line emission before (top), and after (bottom), the scattered emission removal routine is preformed. It is clear that the after the scattered light is subtracted, there is a steeper intensity gradient associated with the intrinsic H$_2$ emission, which is most readily observable in the closeness of contour lines surrounding bright H$_2$ emission, such as the features labelled \textit{A1} \& \textit{A2}. Nebulous emission close the apex of the cavity has also been diminished after scattered light subtraction. The bottom panel also displays, with a red rectangular region, the position and width of the pseudo-slit used to extract the transverse PV diagram seen in Fig. \ref{pvtransverse}. 

%\clearpage

\FloatBarrier

\section{Supplementary spectra}

\label{AppSect: Supp. Spectra}

\begin{figure*}
    \centering
    \includegraphics[width=15cm, trim= 0cm 0cm 0cm 0cm, clip=true]{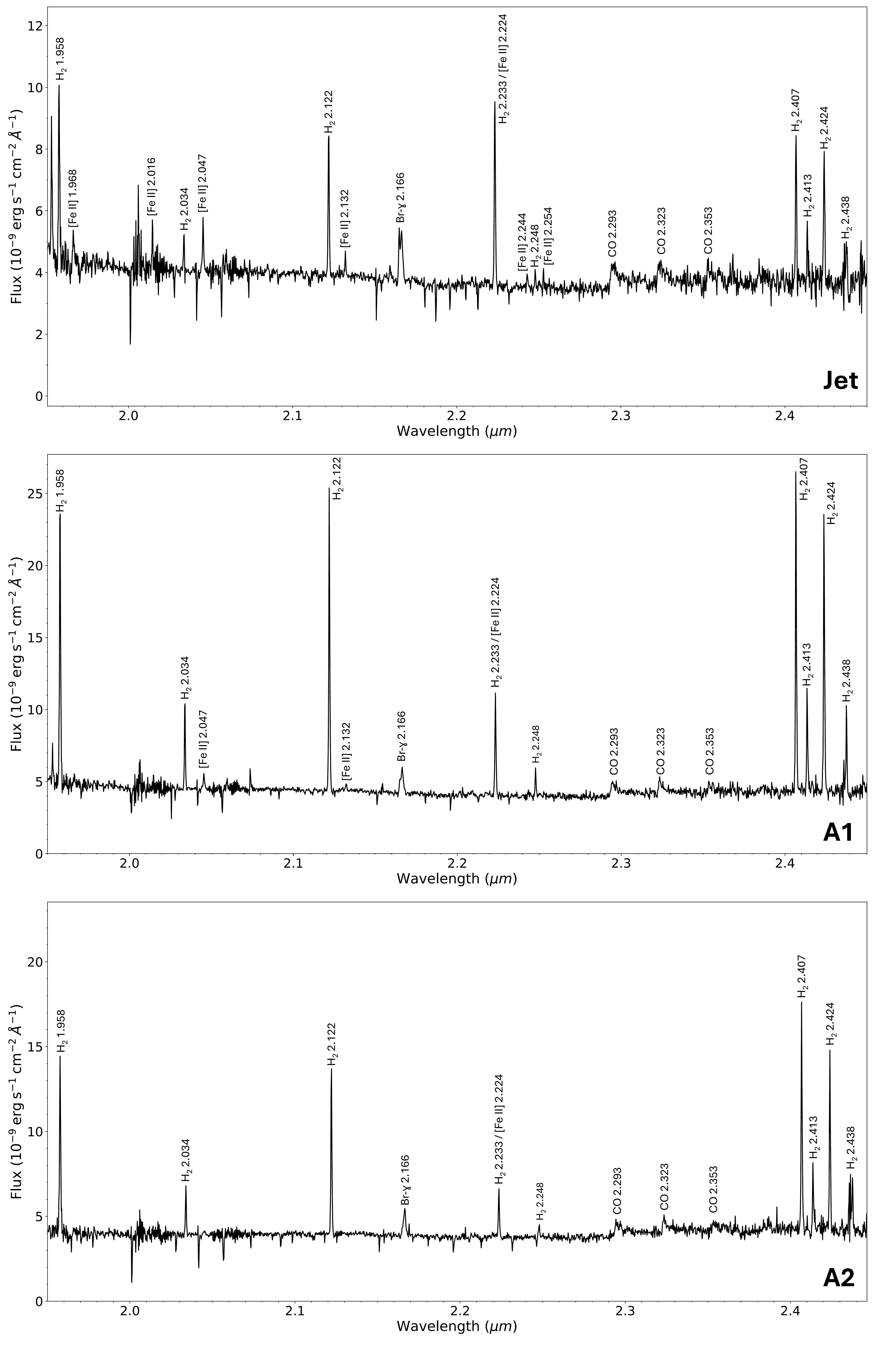}
    \caption{Spectra extracted from the atomic jet knot and the H$_2$ features labelled \textit{A1} \& \textit{A2}. Top: Integrated spectrum extracted using a circular aperture of radius 0\farcs4 from atomic jet knot located at 1\arcsec~offset from source position. Centre: Integrated spectrum extracted using a circular aperture of radius 0\farcs4  from base of left cavity wall traced in H$_2$, position \textit{A1}. Bottom: Integrated spectrum extracted using a circular aperture of radius 0\farcs4 from base of cavity wall, position \textit{A2}.}

    \label{HH4647spectra}
\end{figure*}

Figure \ref{HH4647spectra} displays three integrated spectra extracted from various spatial positions on the object. Molecular hydrogen emission lines (H$_2$), forbidden iron emission ([\ion{Fe}{ii}]), Br-$\gamma$ emission and CO emission are common to all spectra. 

The top spectrum in Fig. \ref{HH4647spectra} was extracted using a circular aperture of radius 0.\farcs4 from the atomic jet knot located approximately 450 au, or 1\farcs0, from the source position. This knot allowed us to study the jet close to the base of the outflow. Atomic forbidden emission lines are identified in this spectrum tracing the collimated jet emanating from the source. Br-$\gamma$ is also identified here with a double peaked structure displaying strong emission in the blueshifted peak. This implies that Br-$\gamma$ here is tracing not only accretion processes onto the star but also ejection, namely the base of the atomic jet. Molecular hydrogen (H$_2$) emission lines also tracing the outflow are identified.  

The central spectrum in Fig. \ref{HH4647spectra} displays a spectrum extracted using a circular aperture of radius 0\farcs4 from the H$_2$ knot labelled \textit{A1} on the left side of the limb-brightened V-shaped cavity traced by molecular hydrogen. The H$_2$ emission lines here are very bright. Comparatively, the spectrum in the bottom of Fig. \ref{HH4647spectra} extracted using a circular aperture of radius 0\farcs4, but from the H$_2$ knot labelled \textit{A2} on the right side of the V-shaped cavity traced by molecular hydrogen. The H$_2$ emission lines here are not as strong as they appear on the left side of the V-shaped cavity. This is expected, as the left knot of the cavity has a significantly greater intensity when compared with the knot on the right side. 

%\section{}
%\label{AppSect: Scattered emission removal}

\FloatBarrier

\end{appendix}
\end{document}